\documentclass[twocolumn]{autart}    % Enable this line and disable the 
\usepackage[utf8]{inputenc} % added by arXiv

\usepackage[numbers]{natbib}
\usepackage{graphicx}
\usepackage{parskip}
\usepackage{amsmath}
\usepackage{amssymb}
\usepackage{epstopdf}
\usepackage{mathtools}
\usepackage{letltxmacro}
\usepackage[table,xcdraw]{xcolor}
\usepackage{hhline} % za duplu liniju u tabel

\newcommand{\nref}[1]{(\ref{#1})}
\newcommand{\bfs}[1]{\boldsymbol{#1}}
\newcommand{\col}[1]{\operatorname{col}\left({#1}\right)}
\newcommand{\diag}[1]{\operatorname{diag}\left({#1}\right)}
\newcommand{\kraj}{\hspace*{\fill} \qed}

\newcommand{\z}[1]{\left( #1 \right)}
\newcommand{\m}[1]{\begin{bmatrix} #1 \end{bmatrix}}
\newcommand{\n}[1]{\left\| #1 \right\|}

\newcommand{\R}{\mathbb{R}}

\newcommand{\proj}{\mathrm{proj}}

\newcommand{\argmin}{\mathrm{argmin}}

\newtheorem{stan_assum}{Standing Assumption}

\begin{document}

\begin{frontmatter}

\title{Learning generalized Nash equilibria in multi-agent dynamical systems via extremum seeking control} % Title, preferably not more 

\thanks[footnoteinfo]{This work was partially supported by the ERC under research project COSMOS (802348). E-mail addresses: \{s.krilasevic-1, s.grammatico\}@tudelft.nl.}

\author{Suad Krilašević} and
\author{\ Sergio Grammatico} 
          
\address{Delft Center for Systems and Control, TU Delft, The Netherlands}

                             % keyword list or with the 
                                          % help of the Automatica 
                                          % keyword wizard

\begin{abstract}                          % Abstract of not more than 200 words.
In this paper, we consider the problem of learning a generalized Nash equilibrium (GNE) in strongly monotone games. First, we propose a novel continuous-time solution algorithm that uses regular projections and first-order information. As second main contribution, we design a data-driven variant of the former algorithm where each agent estimates their individual pseudo-gradient via zero-order information, namely, measurements of their individual cost function values, as typical of extremum seeking control. Third, we generalize our setup and results for multi-agent systems with nonlinear dynamics. Finally, we apply our algorithms to connectivity control in robotic sensor networks and distributed wind farm optimization. 
\end{abstract}

\end{frontmatter}

\section{Introduction}

Multi-agent optimization problems and games with self-interested decision makers or agents appear in many engineering applications, such as demand-side management in smart grids \cite{mohsenian2010autonomous}, \cite{saad2012game}, charging/discharging coordination for plug-in electric vehicles \cite{ma2011decentralized}, \cite{grammatico2017dynamic}, thermostatically controlled loads \cite{li2015market}, \cite{li2015market2}, \cite{grammatico2015mean} and robotic formation control \cite{lin2014distributed}. Typically, in these games, the cost functions and the constraints of the agents are coupled together, e.g. due to common congestion penalties and shared resource capacity, respectively. Since the agents are self-interested, their interaction might be unstable. Thus, one main research area is that of finding (seeking) agent decisions that are self-enforceable, e.g. decisions such that no agent has an incentive to deviate from - the so-called Generalized Nash equilibrium (GNE) \cite{facchinei2010generalized}. From a control-theoretic perspective, in the presence of dynamical agents, the main challenge is to design distributed, possibly almost decentralized, control laws that ensure both the convergence of the agent decisions to a GNE and the asymptotic stability of the equilibrium for the coupled agent dynamics.\\ \\

\emph{Literature review:} The literature on generalized Nash equilibrium problems (GNEPs) is vast - see \cite{facchinei2010generalized} for a survey. Traditional GNE seeking approaches assume that the agents have complete game information, which requires the exchange of large amounts of data over a communication network. Thus, one prominent research focus in recent years has been on semi-decentralized algorithms in which the agents have very limited game information. The class of aggregative games is naturally appealed to such a scenario, since each agent has a cost function which depends on his own decision and on an aggregate (e.g. average) of the decisions of the other agents. The GNEP in aggregative games can be efficiently solved via semi-decentralized algorithms where the aggregate variable is broadcasted by a central coordinator \cite{belgioioso2017semi}, \cite{belgioioso2018douglas}, or via distributed algorithms where the agents locally estimate the aggregate variable  \cite{koshal2016distributed}, \cite{gadjov2019distributed}, \cite{belgioioso2020distributed}. Continuous-time GNE seeking algorithm have been recently proposed as well, all with projections onto the state-dependent tangent cone of the local constraints \cite{de2019continuous}, \cite{mattia2019dynamic}.\\ \\

%%% SG: INstead of [Yi, Pavel 2019], we should put [Gadjov, Pavel, CDC 2019] and [Belgiosios, Nedich, SG, TAC 2020] that are on aggregative games.
%%%By regularizing the operators in a forward-backward algorithm, authors in \cite{yi2019operator} solve the problem of causal computability and present a way to solve, in a distributed way, a wider class of games for both discrete algorithms with projections onto convex sets, and continuous algorithms with projections onto tangents cones \cite{mattia2019dynamic}.\\ \\

We note that in most of the literature, GNE seeking algorithms are for static agents, i.e., where the agent costs instantaneously reflect the chosen decisions. However, this is not the case when the cost functions depend on the internal states of the agents and not on their decisions (control inputs). Let us refer to this class of agents as dynamical agents. The two main approaches to reach a GNE for dynamical agents are passivity-based \textit{first-order} algorithms and payoff-based \textit{zero-order} algorithms. By using a passivity property, in \cite{gadjov2018passivity}, \cite{romano2019dynamic}, Pavel and co-authors design a control law that guarantees convergence to a Nash equilibrium (NE) in a multi-agent system with single and multi-integrator dynamics over a time-invariant network. With the same goal, in \cite{de2019distributed}, De Persis and Grammatico relax the network connectivity assumption in \cite{gadjov2018passivity} by designing a network weight adaptation scheme. In \cite{mattia2019dynamic}, the authors extend the convergence results to GNEPs for the first time. In payoff-based algorithms, each agent can only measure the value of their own cost function, but does not know its analytic form. Most of such algorithms are designed for NEPs with static agents with finite action spaces, e.g. \cite{goto2012payoff}, \cite{marden2009cooperative}, \cite{marden2012revisiting}. In the case of continuous action spaces, the measurements of the cost functions are often used to estimate the pseudo-gradients. Perhaps the most popular class of control algorithms that exploits this principle is that of extremum seeking control (ESC). The main idea is to use perturbation signals to ``excite'' the cost function and estimate its gradient. Since the first general proof of convergence \cite{krstic2000stability}, there has been a strong research effort to extend the original ESC approach \cite{tan2006non}, \cite{ghaffari2012multivariable}, as well as to conceive diverse variants, e.g. \cite{durr2013lie} and \cite{guay2015time}. ESC was used for NE seeking in \cite{frihauf2011nash} where the proposed ESC algorithm is proven to converge to a neighborhood of a NE for nonlinear dynamical agents. The results are extended in \cite{liu2011stochastic} to include stochastic perturbation signals. In \cite{poveda2017framework}, Poveda and Teel propose a framework for the synthesis of a hybrid controller which could also be used for NEPs with nonlinear dynamical agents. 
Unfortunately, all the available ESC algorithms cannot be used for GNE learning/seeking.
The main technical issue is the primal-dual Lagrangian reformulation of the GNEP, which is necessary to decouple the coupling constraints, does not preserve the strong monotonicity of the (extended) pseudo-gradient, the usual background assumption in the literature. This is an important technical challenge and in fact there is still no research on data-driven (zero-order) GNE learning in monotone games for nonlinear dynamical agents.\\ \\

\emph{Contribution}: Motivated by the above literature and open research problem, to the best of knowledge, we consider and solve for the first time the problem of learning a GNE in monotone games with nonlinear dynamical agents. Specifically, our main technical contributions are summarized next:
%with dynamic nonlinear agents which only have access to current measurements of the cost function and are in limited bidirectional communication with a central coordinator. We propose
\begin{itemize}
    \item We design a novel, full-information, semi-decentralized continuous-time GNE seeking algorithm, which uses projections onto fixed convex sets instead of projections onto state-dependent tangent cones as in \cite{mattia2019dynamic}. In this way, the state flow is Lipschitz continuous and admits solutions in the classical sense. We overcome the lack of strong monotonicity of the primal-dual pseudo-gradient thanks to a suitable preconditioning of the operators defining the optimality conditions. 
    \item We design an extremum seeking scheme that learns GNEs in (strongly) monotone games with static agents who perform local computations and communicate with a central coordinator only. Differently from \cite{guay2015time}, where the authors consider an optimization problem, we consider a noncooperative game. Furthermore, we prove that, with a time-scale separation, our algorithm learns GNEs in (strongly) monotone games with nonlinear dynamical agents.  
\end{itemize}
We also apply for the first time semi-decentralized GNE learning to the robot connectivity problem and to wind farm optimization.\\ \\

\emph{Notation}: $\mathbb{R}$ denotes the set of real numbers. For a matrix $A \in \mathbb{R}^{n \times m}$, $A^\top$ and $\|A\|$ denote its transpose and maximum singular value respectively. For vectors $x, y \in \mathbb{R}^{n}$, $x^\top y$ and $\|x \|$ denote the Euclidean inner product and norm, respectively. We denote the unit ball set as $\mathbb{B} \coloneqq \{x \in \R^N\ |\ \n{x} \leq 1\}$. Given $N$ vectors $x_1, \dots, x_N$, possibly of different dimensions, $\col{x_1, \dots x_N} \coloneqq \left[ x_1^\top, \dots, x_N^\top \right]^\top $. Collective vectors are defined as $\bfs{x} \coloneqq \col{x_1, \dots, x_N}$ and for each $i = 1, \dots, N$, $\bfs{x}_{-i} \coloneqq \col{ x_1, \dots,  x_{i -1},  x_{i + 1}, \dots, x_N } $. Given $N$ matrices $A_1$, $A_2$, \dots, $A_N$, $\operatorname{diag}\left(A_{1}, \ldots, A_{N}\right)$ denotes the block diagonal matrix with $A_i$ on its diagonal. For a function $v: \mathbb{R}^{n} \times \mathbb{R}^{m}  \rightarrow \mathbb{R}$ differentiable in the first argument, we denote the partial gradient vector as $\nabla_x v(x, y) \coloneqq \left[\frac{\partial v(x, y)}{\partial x_{1}}^\top, \ldots, \frac{\partial v(x, y)}{\partial x_{N}}^\top\right]^\top \in \mathbb{R}^{n}$. Maximal and minimal eigenvalues of matrix $A$ are denoted as $\sigma_{\textup{max}}(A)$ and $\sigma_{\textup{min}}(A)$ respectively. The mapping $\proj_S : \R^n \rightarrow S$ denotes the projection onto a closed convex set $S$, i.e., $\proj_S(v) \coloneqq \argmin_{y \in S}\n{y - v}$.  The set-valued mapping $\text{N}_{S}: \R^{n} \rightrightarrows \R^{n}$ denotes the normal cone operator for the set $S \subseteq \R^{n}$, i.e., $\text{N}_{S}(x) = \varnothing$ if $x \notin S$, $\left\{v \in \mathbb{R}^{n} | \sup _{z \in S} v^{\top}(z-x) \leq 0\right\}$ otherwise. $\operatorname{Id}$ is the identity operator. $I_n$ is the identity matrix of dimension $n$ and $ \bfs{0}_n$ is vector column of $n$ zeros. The non-negative orthant is defined as $\{\geq 0\} \coloneqq \{v \in\R^n\ |\ v \geq \bfs{0}_n\}$. For a set $\mathcal{M}\coloneqq\{1, \dots, M\}$ and a vector-valued function $\phi \coloneqq \col{(\phi_i(\cdot))_{i \in \mathcal{M}}}: \R \rightarrow \R^M$, we denote $D^+ \phi(t) \coloneqq \col{(\lim \sup_{h \rightarrow 0^+} \frac{\phi_i(t+h)-\phi_i(t)}{h})_{i \in \mathcal{M}}}$.
% For a locally Lipschitz function $\Psi: \R^{n} \rightarrow \R$ and  $\phi: \R \rightarrow \R^n$  which is the solution of the differential equation $\dot{\phi} = f(\phi, u)$, $u \in \R^m$, the upper directional Dini derivative of $\Psi$ in respect to $\phi$ is defined as $D^{+} \Psi(\phi, u) \coloneqq \limsup _{h \rightarrow 0^{+}} \frac{\Psi(\phi+h f(\phi, u))-\Psi(\phi)}{h}$.

\section{Multi-agent dynamical systems}
We consider a multi-agent system with $N$ agents indexed by $i \in \mathcal{I} = \{1, 2, \dots , N \}$, each with the following dynamics:
\begin{subequations}{}
\begin{align}
    \dot{x}_i &= f_i(x_i, u_i), \label{sistem_i}\\
    y_i &= h_i(x_i, \bfs{x}_{-i}), \label{cost_i}
\end{align}{}
\end{subequations}
where $x_i \in \mathcal{X}_i \subset \mathbb{R}^{n_i}$ is the state variable, $u_i \in \Omega_i \in \mathbb{R}^{m_i}$ is the control input (decision variable),  $y_i \in \mathbb{R}$ is the output variable which evaluates the cost function $h_{i}: \mathbb{R}^{n_i} \times \mathbb{R}^{n_{-i}} \rightarrow \mathbb{R}$, and $f_{i}: \mathcal{X}_i \rightarrow \mathbb{R}^{n_i}$ is the state flow mapping.  Let us also define $n \coloneqq \sum n_i$ and $n_{-i} \coloneqq \sum_{j \neq i} n_j$.\\ \\
% We consider a multi-agent system with $N$ agents indexed by $\mathcal{I} = \{1, 2, \dots , N \}$, each with following cost function:
% \begin{align}
%     l_i(\bfs{u}) = J_i(u_i, \bfs{u}_{-i}), \label{eq: cost_static}
% \end{align}{}
% where $u_i \in \mathbb{R}^{m_i}$ is the decision variable,  $l_i \in \mathbb{R}$ is the output variable which evaluates the cost function $J_{i}: \mathbb{R}^{m_i} \times \mathbb{R}^{m_{-i}} \rightarrow \mathbb{R}$. Let us also define $m \coloneqq \sum m_i$ and $m_{-i} \coloneqq \sum_{j \neq i} m_j$.\\ \\
%The agents do not know the analytic expression of their cost function and can only measure it. They also cannot observe the actions of other players.\\ \\
To ensure the existence and uniqueness of the solutions to the state equations, we make a common assumption in the nonlinear system literature \cite[Thm. 3.3]{khalil2002nonlinear}:
\begin{assum}[Existence and uniqueness]\label{assum: exist}
For each $i \in \mathcal{I}$, $f_i$ is locally Lipschitz continuous on $\mathcal{X}_i \times \Omega_i$, where $\mathcal{X}_i$ is a compact set such that any solution to \nref{sistem_i} with $x_i(t_0) \in \mathcal{X}_i$, lies entirely in $\mathcal{X}_i$. \kraj
%there be a compact set $\mathcal{S}_i \subset \mathcal{X}_i$ such that any solution to \nref{sistem_i} with $x_i(t_0) \in \mathcal{S}_i$, lies entirely in $\mathcal{S}_i$. \kraj
\end{assum}{}

Furthermore, we assume that the decision variables of the agents are subject to local constraints $u_i \in \Omega_i$ and coupling constraints $A\bfs{u} \leq b$, where $A \in \R^{q \times m}$, $b \in \R^{q}$, and $\bfs{u} \coloneqq \col{(u_i)_{i \in \mathcal{I}}}$ collects all the control inputs. Let us denote the collection of local constraints as
\begin{align}
    &\bfs{\Omega} \coloneqq\Omega_1 \times \dots \times \Omega_N. 
\end{align}
As the the decision variables are also coupled together, the overall feasible decision set $\bfs{\mathcal{U}}$ is contained in $\bfs{\Omega}$, i.e.
\begin{align}
    &\bfs{\mathcal{U}}\coloneqq\bfs{\Omega} \cap\left\{\bfs{u} \in \mathbb{R}^{m}\ |\ A \bfs{u} \leq b\right\}, \label{def: U}
\end{align}{}%
Let us also denote the feasible set of each agent $i$ as
\begin{align}
    &{\mathcal{U}}_i(u_{-i})\coloneqq{\Omega}_i \cap\left\{{u}_i \in \mathbb{R}^{m_i}\ |\ A \bfs{u} \leq b\right\}.
\end{align}
A common assumption amongst the extremum seeking literature (for example \cite[Ass. 2.1]{krstic2000stability}, \cite[Equ. 3]{guay2017proportional}, \cite[Ass. 2]{poveda2017framework}) is the existence of the steady-state mapping. 
\begin{stan_assum}[Steady-state mapping]\label{assum: steady_state}
For each $i \in \mathcal{I}$, there is a differentiable mapping $\pi_i: \R^{m_i}\rightarrow\R^{n_i}$ (called the steady-state mapping) such that for every $\overline{u}_i \in {\Omega}_i$, it holds that $f_i(\pi_i(\overline{u}_i), \overline{u}_i) = 0$. \kraj
\end{stan_assum}{}
By using the previous definition, let us also define the collective steady-state mappings
% \begin{align}
% &\pi(\bfs{u}) \coloneqq \left[ \begin{array}{c}{\pi_1(u_1)} \\ \vdots \\ {\pi_N(u_N)}
% \end{array}   \right], & \pi_{-i}(\bfs{u}_{-i}) \coloneqq \left[ \begin{array}{c}{\pi_1(u_1)} \\ \vdots \\ {\pi_{i-1}(u_{i-1})} \\ {\pi_{i + 1}(u_{i + 1})} \\ \vdots \\{\pi_N(u_N)}
% \end{array}   \right]. \label{def: steady_state_map}
% \end{align}

\begin{align}
\pi(\bfs{u}) &\coloneqq \col{\z{\pi_i(u_i)}_{i \in \mathcal{I}}}, \nonumber\\
\pi_{-i}(\bfs{u}_{-i}) &\coloneqq \col{\z{\pi_j(u_j)}_{j \in \mathcal{I} \setminus \{i\}}}. \label{def: steady_state_map}
\end{align}

In this paper, we assume that the goal of each agent is to minimize its own steady-state  cost function, i.e.,
\begin{align}
\forall i \in \mathcal{I}:\ \min_{u_i \in {\mathcal{U}_i}(\bfs{u}_{-i})}  J_i(u_i, \boldsymbol{u}_{-i}), \label{def: dyn_game} 
\end{align}{}
where 
\begin{align}
    J_i(u_i, \boldsymbol{u}_{-i}) \coloneqq h_i(\pi(u_i), \pi_{-i}(\bfs{u}_{-i})), \label{eq: cost}
\end{align}{}
% We assume that the goal of each agent is to minimize its own cost function, i.e.,
% \begin{align}
% \forall i \in \mathcal{I}:\ \min_{\bfs{u} \in {\mathcal{U}_i}} J_i(u_i, \bfs{u}_{-i}), \label{goal}
% \end{align}{}
which depends on decision variables of other agents as well. From a game-theoretic perspective, we consider the problem to compute a generalized Nash equilibrium (GNE), as formalized next.
\begin{defn}[Generalized Nash equilibrium]\hfill\quad
A set of control actions $\bfs{u}^*\coloneqq\col{u_i^*}_{i \in \mathcal{I}}$ is a generalized Nash equilibrium if, for all $i \in \mathcal{I}$,
\begin{align}
    u_{i}^{*} \in \underset{v_{i} \in \Omega_i}{\operatorname{argmin}}\ J_{i}\left(v_{i}, \bfs{u}_{-i}^{*}\right) \mathrm{ s.t. }\left(v_{i}, \bfs{u}_{-i}^{*}\right) \in \bfs{\mathcal{U}}.  \label{def: gne}
\end{align}\end{defn}{}
with $\bfs{\mathcal{U}}$ as in \nref{def: U} and $J_i$ as in \nref{eq: cost}.\kraj \\ \\
In plain words, a set of inputs is a GNE if no agent can improve its steady-state cost function by unilaterally changing its input. To ensure the existence of the GNE, we postulate the following basic assumption \cite[Thm. 2]{facchinei2007generalized}:
\begin{stan_assum}[Regularity] \label{sassum: regularity}
For each $i \in \mathcal{I}$, the function $J_i$ in \nref{eq: cost} is continuous; the function $J_{i}\left(\cdot, \bfs{u}_{-i}\right)$ is convex for every $\bfs{u}_{-i}$. For each $i \in \mathcal{I}$, the set $\Omega_i$ is non-empty, closed and convex; $\bfs{\mathcal{U}}$ is non-empty and satisfies Slater's constraint qualification. \kraj
\end{stan_assum}{}
More precisely, in this paper we focus on a subclass of GNE called variational GNE (v-GNE) \cite[Def. 3.11]{facchinei2007generalized}. A collective decision $\bfs{u}^*$ is a v-GNE in \nref{def: gne} if and only if there exists a dual variable $\lambda \in \R^q$ such that the following KKT conditions are satisfied \cite[Th. 4.8]{facchinei2007generalized}:

\begin{subequations}
\begin{align}
\mathbf{0}_{m} &\in F\left(x^{*}\right)+A^{\top} \lambda^{*}+\mathrm{N}_{\bfs{\Omega}}\left(x^{*}\right) \label{equ: kkt1}\\
\mathbf{0}_{q} &\in-\left(A x^{*}-b\right)+\mathrm{N}_{\geq 0}^{q}\left(\lambda^{*}\right). \label{equ: kkt2}
\end{align}{}
\end{subequations}{}

By stacking the partial gradients $\nabla_{u_i} J_i(u_i, \boldsymbol{u}_{-i})$ in to single vector, we form the pseudo-gradient mapping
\begin{align}
    F(\boldsymbol{u}):=\operatorname{col}\left(\left(\nabla_{u_{i}} J_{i}\left(u_{i}, \bfs{u}_{-i}\right)\right)_{i \in \mathcal{I}}\right). \label{eq: pseudogradient}
\end{align}
We assume that there exists a central coordinator who is capable of bidirectional communication on a star-shaped network with the agents, which is a frequent assumption in semi-decentralized algorithms \cite{belgioioso2018douglas}, \cite{belgioioso2019distributed}. The central coordinator is tasked with computation of the dual variables. For ease of notation, we index the coordinator with $0$ and define $\mathcal{I}_0 \coloneqq \mathcal{I} \cup \{0\}$.\\

Let us postulate additional common assumptions (\cite[Std. Ass. 2]{de2019continuous}, \cite[Ass. 1]{de2019distributed}) in order to assure the convergence of the algorithm we propose latter on.
\begin{stan_assum}[Well-behavedness]
For each $i \in \mathcal{I}$, $J_i$ in \nref{eq: cost} is twice differentiable, Lipschitz continuous, and its gradient $\nabla J_i$ is $\ell$-Lipschitz continuous, with $\ell > 0$. The pseudo-gradient mapping $F$ in \nref{eq: pseudogradient} is $\mu$-strongly monotone, i.e., for any pair $\bfs{u}, \bfs{v} \in \mathbb{R}^n$, $(\bfs{u} - \bfs{v})^\top (F(\bfs{u}) - F(\bfs{v})) \geq \mu \|\bfs{u} - \bfs{v} \|$, with $\mu > 0$. \kraj
\end{stan_assum}{}
% \begin{stan_assum}
% The pseudo-gradient mapping is strongly monotone and Lipschitz continuous:  and $\|F(\bfs{u}) - F(v) \| \leq L \| \bfs{u} - v\|$, for some $\mu > 0$ and $L > 0$. \kraj
% \end{stan_assum}{}

Another common assumption is (local) exponential stability of the equilibrium points $\pi_i(u_i)$, under constant input ($\dot{u}_i = 0$) \cite[Ass. 2.2]{krstic2000stability}, \cite[Ass. 4.2]{frihauf2011nash}. Thus, with the change of coordinates $z_i \coloneqq x_i - \pi_i(u_i)$, we also adopt the following assumption throughout the paper:

\begin{stan_assum}[Lyapunov stability]\label{assum: lyap}
For each $i \in \mathcal{I}$, there is a smooth Lyapunov function, $z_i \mapsto V_i(z_i, u_i)$, with Lipschitz continuous partial derivatives, i.e. for every constant $\overline{u}_i \in \mathcal{U}_i$, it holds that
\begin{subequations}
\begin{align}
    \underline{\alpha}_i \|z_i\|^2 \leq V_i(z_i, \overline{u}_i) &\leq \overline{\alpha}_i \|z_i \|^2 \\
    \frac{\partial V_i}{\partial z_i} (z_i, \overline{u}_i)^\top f(z_i + \pi_i(\overline{u}_i), \overline{u}_i) &\leq -\kappa_i \|z_i\|^2 \\
    \frac{\partial V_i}{\partial z_i}(0, \overline{u}_i) &= 0
\end{align}{}
\end{subequations}{}
for some positive constants $\underline{\alpha}_i$, $\overline{\alpha}_i$ and $\kappa_i$. Moreover, for every constant $\overline{u}_i \in \mathcal{U}_i$, it holds that
\begin{align}
    & &\frac{\partial V_i}{\partial {u}_i} (0, \overline{u}_i) = 0. &\kraj
\end{align}
\end{stan_assum}{}
\section{Generalized Nash Equilibrium seeking for static agents}
Let us start from the case of static agents to highlight the proposed algorithm and its integration with the zero-oder gradient scheme. 
\begin{assum}[Static agents]\label{assum: static agents}\hfill \\
For each $i \in \mathcal{I}$, $x_i = u_i$ (in place of \nref{sistem_i}). \kraj
\end{assum}
We propose two control schemes for GNE seeking with static agents. In the first, the agents have perfect information about the decisions of other agents and know the analytic expression of their partial gradient. The second scheme is data-driven, i.e. the agents have access to the output of their own cost function only.
\subsection{Gradient-based case}
In our first GNE seeking algorithm, each agent updates their decision, $u_i$, based on decisions of all other agents and the dual variable. A central coordinator updates the dual variable and broadcasts it amongst the agents:
\begin{align}
\forall i \in \mathcal{I}: \dot{u}_i &= -u_i + \proj_{\Omega_i}\z{u_i - \gamma_i(\nabla_{u_i}J_i(\bfs{u}) + A_i^\top\lambda)} \nonumber\\
\dot{\lambda} &= - \lambda + \operatorname{proj}_{\geq \boldsymbol{0}}(\lambda + \gamma_0 (A\bfs{u} - b + 2A \dot{\bfs{u}})),\nonumber
\end{align}{}
or in collective form
\begin{align}
\m{\dot{\bfs{u}} \\ \dot{\lambda}}&= - \m{\bfs{u} \\ \lambda} + \m{\operatorname{proj}_{\bfs{\Omega}}\left( \bfs{u} - \Gamma (F(\bfs{u}) + A^\top\lambda) \right) \\ \operatorname{proj}_{\geq \boldsymbol{0}}(\lambda + \gamma_0 (A\bfs{u} - b + 2A \dot{\bfs{u}}))}, \label{eg: control full info}
\end{align}{}
where $\lambda \in \R^{q}$, $(\gamma_i)_{\ i \in \mathcal{I}}$ are the step sizes chosen by the agents; $\Gamma = \diag{(\gamma_i I_{m_i})_{i \in \mathcal{I}}}$ and $\gamma_0$ is the step size chosen by the central coordinator. Now we state our first theorem:

\begin{thm}[v-GNE seeking]\hfill\\ \label{thm: full_info}
Let the Standing assumptions and Assumption \ref{assum: static agents} hold and let $(\bfs{u}(t),\lambda(t))_{t \geq 0}$ be the solution to  \nref{eg: control full info}. Then, there exist small enough $(\gamma_i)_{i \in \mathcal{I}_0}$ such that the pair $(\bfs{u}(t), \lambda(t))_{t \geq 0}$ converges to some $(\bfs{u}^*, \lambda^*)\in \bfs{\mathcal{U}} \times \R_{\geq 0}$, where $\bfs{u}^*$ is the variational generalized Nash equilibrium of the game in \nref{def: gne}. \kraj
\end{thm}{}

\begin{pf}
See Appendix A.\hfill $\blacksquare$
\end{pf}{}

\begin{rem}
The algorithm in \nref{eg: control full info} is semi-decentralized, meaning that the update and the broadcast of the dual variable is managed by a central coordinator. A fully distributed variant, where each agent has their own copy of the dual variable and their consensus is imposed, can be derived similarly to \cite[Sec. 3]{mattia2019dynamic}.
% a novel continuous-time preconditioned variant of the forward-backward algorithm \cite{abbas2015dynamical}. Unlike the gradient based algorithms in \cite{de2019continuous} and \cite{mattia2019dynamic}, ours uses projections onto convex sets instead of projections onto tangent cones. In this way, the flow map is Lipschitz continuous, which is required in the case for the parameter estimation scheme.
\end{rem}

\subsection{Data-driven case}
In the limited information case, we consider that the agents have access to the cost output only. We emphasize that in this case, they neither know the actions of the other agents, nor they know the analytic expressions of their partial gradients. In fact, this is a standard setup used in extremum seeking (\cite{krstic2000stability}, \cite{guay2017proportional}, \cite{poveda2017framework} among others). However, the agents can communicate with a central coordinator, to whom they send their decision variable and its derivative.\\

Let us first evaluate the time derivative of the cost output $l_i = J_i(u_i, \bfs{u}_{-i})$ along the trajectories of $\bfs{u}$:
\begin{align}
    %\dot{l}_i &= \nabla_{\bfs{u}_{-i}}J(u_i, \bfs{u}_{-i})^\top \dot{\bfs{u}}_{-i} + \nabla_{u_i}J(u_i, \bfs{u}_{-i})^\top \dot{u}_i 
    \dot{l}_i &=\theta_i^0(\bfs{u})   + \theta_i^1(\bfs{u})  \dot{u}_{i} = [1, \dot{u}_i^\top]\theta_i(\bfs{u}), \label{parametrizacija y}
    %&= \theta_i^0  \dot{\bfs{u}}_{-i}  + \theta_i^1  \dot{u}_{i},
\end{align}{}
where we define
\begin{align}
    \theta_i^0 = \theta_i^0(\bfs{u})& \coloneqq \nabla_{\bfs{u}_{-i}}J_i(u_i, \bfs{u}_{-i})^\top \dot{\bfs{u}}_{-i} \\
    \theta_i^1 = \theta_i^1(\bfs{u})&\coloneqq \nabla_{u_i}J_i(u_i, \bfs{u}_{-i})^\top. \\
    \theta_i = \theta_i(\bfs{u})&\coloneqq [\theta_{i}^{0}, \theta_{i}^{1\top}]
\end{align}{}

The variable $\theta_{i}^{0}$ measures the influence of the decision variables of the other agents  on the cost output of agent $i$. The variable $\theta_{i}^{1}$ measures the effect of the decision variable of agent $i$ on the cost output $l_i$, which is needed for \nref{eg: control full info}. To estimate the local $\theta_{i}^{0}$ and $\theta_{i}^{1}$, we use a time-varying parameter estimation approach, as proposed in \cite{guay2017proportional} for centralized optimization. Let us provide a basic intuition first.\\ \\
Let $\hat{l}_i$ and $\hat{\theta}_i$ be estimations of the output $l_i$ and the variable $\theta_i$ respectively and let $e_i = l_i - \hat{l}_i$ be the estimation error. Then, the estimator model in \nref{parametrizacija y} for agent $i$ is given by 
\begin{align}
    \dot{\hat{l}}_i&=[1, \dot{u}_i^\top] \hat{\theta}_i+K_i e_i+c_i^{\top} \dot{\hat{\theta}}_i, \label{estim1}
\end{align}
where $K_i$ is a free design parameter. Note that the first two terms on the right-hand side resemble high gain observer schemes \cite{oh1997nonlinear}. As the structure of the problem does not directly allow the use of high gain observers, it is necessary to introduce some other dynamics into the estimation. This is the primary role of the third term in \nref{estim1}. Therefore, the dynamics of $c_i(t)$ are choosen as 
\begin{align}
\dot{c}_i^{\top}&=-K_i c_i^{\top}+[1, \dot{u}_i^\top]. \label{ci dyn}
\end{align}
Let us also introduce an auxiliary variable $\eta_i = e_i - c_i^\top \tilde{\theta}_i$, with dynamics $D^+{\eta}_i=-K_i \eta_i-c_i^{\top} D^+{\theta}$, and its estimate $\hat{\eta}_i$, with dynamics
\begin{align}{}
\dot{\hat{\eta}}_i&=-K_i \hat{\eta_i}.
\end{align}
%The original parameter estimation law in \cite{adetola2008finite} was designed for constant parameters, therefore $\eta_i = \hat{\eta_i}$. For time-varying parameters, we still want to use $\eta_i$, but the additional term $-c_i^{\top} \dot{\theta}$ does not allow for its calculation, since the rate of change of the parameters is unknown. This is why the estimate $\hat{\eta}_i$ is used in the parameter estimation law. 
It is also necessary to define a symmetric, positive definite matrix variable $\Sigma_i \in \mathbb{R}^{(m_i + 1) \times (m_i + 1)}$ with dynamics given by
\begin{align}
&\dot{\Sigma}_i=c_i c_i^{\top}-\rho_i \Sigma_i+\sigma_i I 
&\Sigma_i(0) = \Sigma_i^0,\label{estim2}
\end{align}{}where $\rho_i$, $\sigma_i$ and $\Sigma_i^0$ are free design parameters. We note that, originally, in \cite{adetola2008finite}, $\dot{\Sigma}_i=c_i c_i^{\top}$, but this proved to be inconvenient in practical implementations, as the elements of $\Sigma_i$ grow unbounded. Instead, as in \nref{estim2}, dynamics of $\Sigma_i$ behave as a first-order system. The third term is added so that the matrix is always invertible. Equations \nref{estim1}-\nref{estim2} form the parameter update law presented in \cite{adetola2008finite}:
\begin{align}
\dot{\hat{\theta}}_i=\operatorname{\Pi}_{\Theta_i}\left(\hat{\theta}_i, \Sigma_i^{-1}(c_i(e_i-\hat{\eta_i})-\sigma_i \hat{\theta}_i)\right), \label{estimlast}
\end{align}{}where $\operatorname{\Pi}_{\Theta_i}(\hat{\theta}, v)$ denotes the projection of the vector $v$ onto the tangent cone of the set $\Theta_i$ at $\hat{\theta}$, as defined by Equation 2.14 in \cite{nagurney2012projected}. This implies that if the starting value $\hat{\theta}_i (0)$ is in $\Theta_i$, so is $\hat{\theta}_i(t) $. The set $\Theta_i$ represents the expected (possible) values of $\theta_i$. Let us also define $\Theta \coloneqq \Theta_1 \times \dots \times \Theta_N$. \\ \\

We are finally ready to propose our semi-decentralized v-GNE learning algorithm:
\begin{align*}
    \forall i \in \mathcal{I}: \dot{u}_i &= -u_i + \proj_{\Omega_i}\z{u_i - \gamma_i(\hat{\theta}_i^1 + A_i^\top\lambda) + d_i}, \\
    \dot{\lambda}& = - \lambda + \operatorname{proj}_{\geq \boldsymbol{0}}(\lambda + \gamma_0 (A\bfs{u} - b + 2A \dot{\bfs{u}})), 
\end{align*}{}
where $A_i^\top$ represents the ith matrix row of $A^\top$, which contains the constraints of agent $i$ and $d_i$ represents the perturbation signal of agent $i$. In collective form, it can be written as
\begin{align}
\m{\dot{\bfs{u}} \\ \dot{\lambda}} = -\m{\bfs{u} \\ \lambda} + \m{\operatorname{proj}_{\bfs{\Omega}}\left( \bfs{u} - \Gamma (\hat{\bfs{\theta}}^1 + A^\top\lambda) + d \right) \\ \operatorname{proj}_{\geq \boldsymbol{0}}(\lambda + \gamma_0 (A\bfs{u} - b + 2A \dot{\bfs{u}}))}, \label{eq: control law}
\end{align}{}
where $\hat{\bfs{\theta}}^1:=\operatorname{col}\Big(\left(\hat{\theta}^1_i\right)_{i \in \mathcal{I}}\Big)$. As in \cite[Ass. 5]{guay2017proportional}, for the parameter estimation scheme to converge, we postulate a persistency of excitation (PE) assumption  for all agents.\\ \\

\begin{assum}[Persistence of excitation]\hfill \\ \label{assum: excite}
For each $i \in \mathcal{I}$, there are positive constants $\alpha_{i}$ and $T_i$ such that 
\begin{align}
&\int_{t}^{t+T_i} c_i(\tau) c_i(\tau)^\top d \tau \geq \alpha_{i} I,
&\text{for all } t>0,
\end{align}{}where $c_i$ is the solution to \nref{ci dyn}.\kraj
\end{assum}
We conclude the section with the convergence result. For any initial condition of the decision variables, there exist gains such that the control variables converge to an arbitrarily small neighborhood of the v-GNE. %Size of the neighborhood is mostly dependent on the amplitude of the perturbations.

\begin{thm}[v-GNE static learning]\hfill \\
% Let the Standing Assumptions and Assumptions \ref{assum: static agents}  and \ref{assum: excite} hold, let $D$ be the largest amplitude of the perturbation signals $\{d_i\}_{i \in \mathcal{I}}$ and $D_\textup{d}$ be the largest amplitude of the time derivative of the perturbation signals $\{\dot{d}_i\}_{i \in \mathcal{I}}$, and let $(\bfs{u}(t), \lambda(t))_{t \geq 0}$ be the closed-loop solution to \nref{estim1} -- \nref{eq: control law}. Then, there exist gains $(K_i, \rho_i, \sigma_i)_{i \in \mathcal{I}}$, $(\gamma_i)_{\ i \in \mathcal{I}}$ and $\nu_0$, such that for any initial condition $\bfs{u}(0) \in \bfs{\mathcal{U}}$, the pair $(\bfs{u}(t), \lambda(t))$ converges to a neighborhood of size $\mathcal{O}(D^2 + {D}_\textup{d}^2 + \n{F(u^*)}^2)$ of some $(\bfs{u}^*, \lambda^*)$, where $\bfs{u}^*$ is a variational generalized Nash equilibrium of the game in \nref{def: gne}. \kraj

Let the Standing Assumptions and Assumptions \ref{assum: static agents}  and \ref{assum: excite} hold and let $(s(t) \coloneqq (\hat{\bfs{\eta}}(t), \hat{\bfs{\theta}}(t),  \bfs{u}(t), \lambda(t)))_{t \geq 0}$ be the closed-loop solution to \nref{estim1}--\nref{estim2}, \nref{eq: control law}. Then, for any compact set $\mathcal{K}$ and any $\varepsilon > 0$, there exist small enough parameters $(\tfrac{1}{K_i}, \tfrac{1}{\rho_i}, \sigma_i, \gamma_i)_{i \in \mathcal{I}}$ and $\gamma_0$, such that for every solution with $s(0) \in \mathcal{K}$, $\bfs{u}(t)$ converges to the set $\{ \bfs{u}^*\} + \varepsilon \mathbb{B}$, where $\bfs{u}^*$ is a variational generalized Nash equilibrium of the game in \nref{def: gne}. \kraj
\end{thm}{}
\begin{pf}
See Appendix B. \hfill $\blacksquare$
\end{pf}{}

\section{Generalized Nash equilibrium learning for dynamical agents}
In this section, we propose a control scheme for generalized Nash equilibrium learning for dynamical agents. We consider a data-driven scenario only, i.e. the agents have access to the cost output measurements and the information that is given to them by a central coordinator. They are not aware of the analytic expression of their steady-state cost function, nor of their pseudo-gradient, nor can they observe the states and decisions of the other agents.\\ \\ 

For the multi-agent dynamical system 
\begin{align}
    \epsilon \m{\dot{x}_1 \\ \vdots \\ \dot{x}_N} = \epsilon \dot{\bfs{x}} = \m{f_1(x_1, u_1) \\ \vdots \\ f_N(x_N, u_N)} =  f(\bfs{x}, \bfs{u}), \label{eq: fast_dynamics}
 \end{align}{}where $\epsilon > 0$ is a time scale separation constant, with the objective of reaching a neighborhood of a v-GNE, we propose the same control law as in \nref{eq: control law}, with the distinction that $\hat{\bfs{\theta}}^1$ is estimated by a parameter estimation scheme \nref{estim1} -- \nref{estimlast}, where we collect the measurements of the output $y_i$ in \nref{cost_i} instead of $J_i(u_i, \bfs{u}_{-i})$ directly. Thus, the estimation error is hereby redefined as 
\begin{align}
    e_i = y_i - \hat{l}_i. \label{def: new_error}
\end{align}{}
We conclude the section with the most general theoretical result of the paper, namely, the convergence of the closed-loop dynamics to a neighborhood of a v-GNE of the game in \nref{def: dyn_game}.

\begin{thm}[{v-GNE dynamic learning}]\hfill \\
% Let the Standing Assumptions and Assumptions \ref{assum: exist} and  \ref{assum: excite} hold, let $D$ be the largest amplitude of the perturbation signals $\{d_i\}_{i \in \mathcal{I}}$, let ${D}_{\textup{d}}$ the largest amplitude of the time derivative of the perturbation signals $\{\dot{d}_i\}_{i \in \mathcal{I}}$, and let ($\bfs{x}(t), \bfs{u}(t), \lambda(t)$) for $t > 0$ be the closed-loop solution to \nref{eq: fast_dynamics}, \nref{def: new_error}, \nref{estim1}--\nref{eq: control law}. Then, there exist gains $(K_i, \rho_i, \sigma_i)_{i \in \mathcal{I}}$, small enough step sizes $(\gamma_i)_{i \in \mathcal{I}}$, $\nu_0$ and a time scale separation constant $\epsilon$, such that for any initial condition $\bfs{u}(0) \in \bfs{\mathcal{U}}$, $\bfs{x}(0) \in \mathcal{X}$,  $(\bfs{x}(t), \bfs{u}(t), \lambda(t))$ converges to a neighborhood of size $\mathcal{O}(D^2 + {D_\textup{d}}^2 + \n{F(u^*)}^2)$ of some $(\tilde{\bfs{x}}, \bfs{u}^*, \lambda^*) = (\pi(\bfs{u}), \bfs{u}^*, \lambda^*)$, where $\bfs{u}^*$ is a variational generalized Nash equilibrium of the game in \nref{def: dyn_game}. \kraj

Let the Standing Assumptions and Assumptions \ref{assum: static agents}  and \ref{assum: excite} hold and let $(s(t) \coloneqq (\hat{\bfs{\eta}}(t), \hat{\bfs{\theta}}(t), \bfs{x}(t),  \bfs{u}(t), \lambda(t)))_{t \geq 0}$ be the closed-loop solution to  \nref{estim1}--\nref{estim2}, \nref{eq: control law}, \nref{eq: fast_dynamics}, \nref{def: new_error}. Then, for any compact set $\mathcal{K}$ and any $\varepsilon > 0$, there exist small enough parameters $(\tfrac{1}{K_i}, \tfrac{1}{\rho_i}, \sigma_i, \gamma_i)_{i \in \mathcal{I}}$, $\gamma_0$ and $\epsilon$ such that every solution with $s(0) \in \mathcal{K}$, $(\bfs{x}(t), \bfs{u}(t))$ converges to an $\varepsilon$ neighborhood of some $( \pi(\bfs{u}(t)),\bfs{u}^*)$, where $\bfs{u}^*$ is a variational generalized Nash equilibrium of the game in \nref{def: gne}. \kraj

\end{thm}{}
\begin{pf}
See Appendix C. \hfill $\blacksquare$
\end{pf}{}

\section{Illustrative applications}
\subsection{Connectivity control in robotic swarms}
The problem of connectivity control has been considered in \cite{stankovic2011distributed} as a Nash equilibrium problem. In many practical scenarios, multi-agent systems, besides their primary objective, are designed to uphold certain connectivity as their secondary objective.  In what follows, we consider a similar problem in which each agent is tasked with finding a source of an unknown signal while maintaining certain connectivity. Unlike \cite{stankovic2011distributed}, we require the existence of a central coordinator and we allow for coupled restrictions on the decisions variables. Additionally, we model the agents as unicycles with setpoint regulators, which does not require a constant angular velocity as in \cite{stankovic2011distributed}.\\ \\

\begin{figure}
    \centering
%% Accompanies image file 'roboti_crtez.pdf' (pdf, eps, ps)
\begingroup%
  \makeatletter%
  \providecommand\color[2][]{%
    \errmessage{(Inkscape) Color is used for the text in Inkscape, but the package 'color.sty' is not loaded}%
    \renewcommand\color[2][]{}%
  }%
  \providecommand\transparent[1]{%
    \errmessage{(Inkscape) Transparency is used (non-zero) for the text in Inkscape, but the package 'transparent.sty' is not loaded}%
    \renewcommand\transparent[1]{}%
  }%
  \providecommand\rotatebox[2]{#2}%
  \newcommand*\fsize{\dimexpr\f@size pt\relax}%
  \newcommand*\lineheight[1]{\fontsize{\fsize}{#1\fsize}\selectfont}%
  \ifx\svgwidth\undefined%
    \setlength{\unitlength}{261.39908887bp}%
    \ifx\svgscale\undefined%
      \relax%
    \else%
      \setlength{\unitlength}{\unitlength * \real{\svgscale}}%
    \fi%
  \else%
    \setlength{\unitlength}{\svgwidth}%
  \fi%
  \global\let\svgwidth\undefined%
  \global\let\svgscale\undefined%
  \makeatother%
  \begin{picture}(1,0.46984068)%
    \lineheight{1}%
    \setlength\tabcolsep{0pt}%
    \put(0,0){\includegraphics[width=\unitlength,page=1]{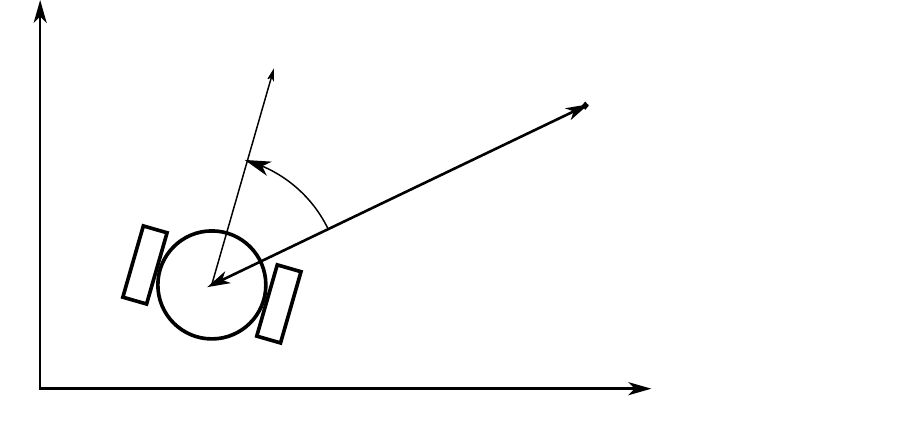}}%
    \put(0.42944733,0.21679547){\color[rgb]{0,0,0}\makebox(0,0)[lt]{\lineheight{1.25}\smash{\begin{tabular}[t]{l}$R_i$\end{tabular}}}}%
    \put(0.31718371,0.28230172){\color[rgb]{0,0,0}\makebox(0,0)[lt]{\lineheight{1.25}\smash{\begin{tabular}[t]{l}$\phi_i$\end{tabular}}}}%
    \put(0.59885335,0.37504289){\color[rgb]{0,0,0}\makebox(0,0)[lt]{\lineheight{1.25}\smash{\begin{tabular}[t]{l}$(u_i^x, u_i^y)$\end{tabular}}}}%
    \put(0.64774311,0.00809197){\color[rgb]{0,0,0}\makebox(0,0)[lt]{\lineheight{1.25}\smash{\begin{tabular}[t]{l}$x$\end{tabular}}}}%
    \put(-0.00347439,0.43344832){\color[rgb]{0,0,0}\makebox(0,0)[lt]{\lineheight{1.25}\smash{\begin{tabular}[t]{l}$y$\end{tabular}}}}%
    \put(0.17596128,0.0654839){\color[rgb]{0,0,0}\makebox(0,0)[lt]{\lineheight{1.25}\smash{\begin{tabular}[t]{l}$(x_i, y_i)$\end{tabular}}}}%
    \put(0,0){\includegraphics[width=\unitlength,page=2]{roboti_crtez.pdf}}%
  \end{picture}%
\endgroup%

    \caption{Variables of the unicycle reference controller.}
    \label{fig:unicycle}
\end{figure}

Consider a multi-agent system consisting of unicycle vehicles, indexed by $i \in \{1, \dots N\}$, with the following dynamics:
\begin{align}
    \dot{x}_i &= v_i \cos{\theta_i}, \nonumber \\
    \dot{y}_i &= v_i \sin{\theta_i}, \nonumber \\
    \dot{\theta}_i &= \omega_i,
\end{align}{}
where $x_i, y_i, \theta_i$ represent the state variables, and $v_i, \omega_i$ represent the control inputs. Each unicycle implements the following feedback controller used for setpoint regulation:
\begin{align}
v_i &=K_{i}^1 \cdot R_i \cdot \cos \phi_i, \nonumber\\
\omega_i &=-K_{i}^2 \cdot \sin (2\phi_i) -K_{i}^2 \cdot \phi_i,
\end{align}
% \begin{align}
%     \dot{r}_i &= - v_i \cos{\theta_i}, \nonumber \\
%     \dot{\phi}_i &= \omega + \dfrac{v_i}{r_i} \sin{\phi_i},
% \end{align}{}
for some $K_{i}^1, K_{i}^2 > 0$, which was studied in \cite{lee2000stable}. A graphical representation of the variables $R_i$ and $\phi_i$ is shown on Figure \ref{fig:unicycle}. Therefore, we rewrite the closed-loop nonlinear dynamics as
\begin{align}
    &\m{\dot{x}_i \\ \dot{y}_i\\ \dot{\phi}_i} = \m{-K_{i}^1 \|r_i - u_i \|\cos \phi_i \cos\z{\phi_i - \arctan \dfrac{y_i}{x_i}} \\ -K_{i}^1 \|r_i - u_i \|\cos \phi_i \sin\z{\phi_i - \arctan \dfrac{y_i}{x_i}}\\-K_{i}^2  \phi_i}, 
\end{align}{}
where $r_i = \col{x_i, y_i}$ and $u_i = \col{u_i^x, u_i^y}$ is the input of the transformed system, which represents the coordinates of the setpoint. For each $i$, the steady-state mapping is then given by $\pi_i(u_i) = \col{u_i, 0}$. \\ \\
Each agent is tasked with locating a source of a unique unknown signal. The strength of all signals abides by the inverse-square law, i.e. proportional to $1/r^2$. Therefore, the inverse of the signal strength can be used in the cost function. Additionally, the agents must not drift apart from each other too much, as they should provide quick assistance to each other in case of critical failure. This is enforced in two ways: by incorporating the signal strength of the fellows agents in the cost functions and by communicating with the central coordinator. Thus, we design the cost output and position constraints as
\begin{align}
    \forall i \in \mathcal{I}: 
    \Bigg\{\begin{array}{l} y_i = \|r_i - r_i^s \|^2 + c\sum_{j \in \mathcal{I}_{-i}}\| r_i - r_j\|^2,\\ 
    \n{\col{(u_i - u_j)_{j \in \mathcal{I}_{-i}}}}_\infty \leq b\end{array}
\end{align}
where $\mathcal{I}_{-i} \coloneqq \mathcal{I}\setminus \{i\}$, $c, b > 0$ and $r_i^s$ represents the position of the source assigned to agent $i$. The safe traversing area is described by a rectangle: $[x_{\textup{min}}, x_{\textup{max}}] \times [y_{\textup{min}}, y_{\textup{max}}]$. \\ \\

\begin{figure}
    \centering
    \includegraphics[width = \linewidth]{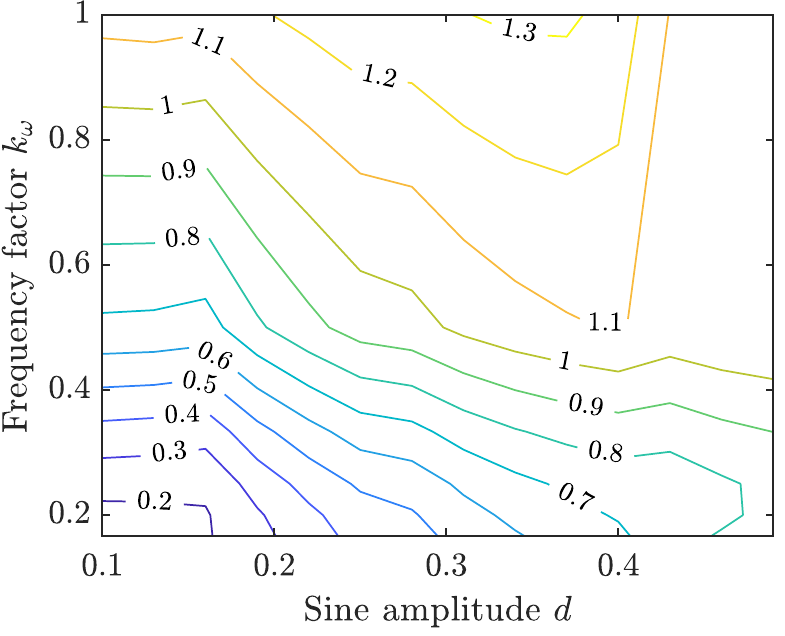}
    \caption{Distance of the final average steady-state trajectory from the v-GNE for agent 4.}
    \label{fig: convergence_distance}
\end{figure}

\begin{figure}[ht]
    \centering
    \includegraphics[width = \linewidth]{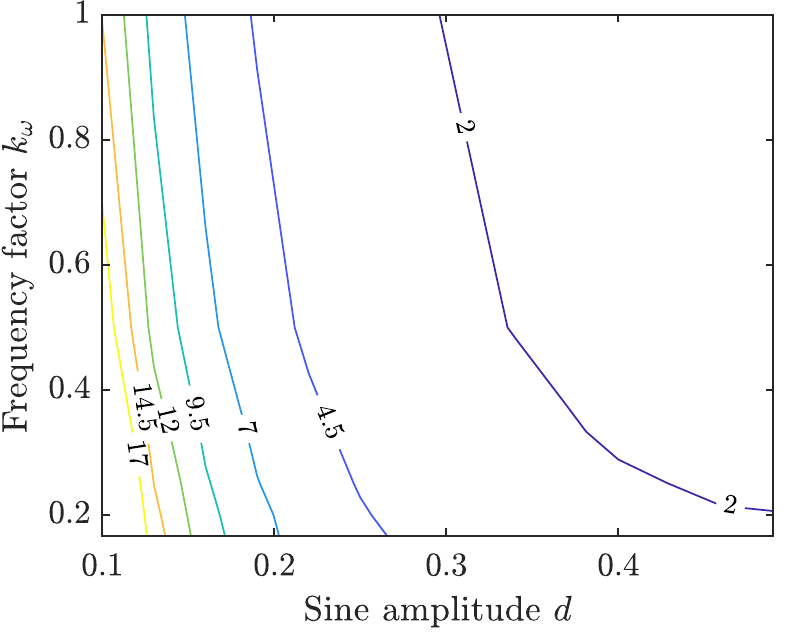}
    \caption{Hours required to enter a ball of size $\varepsilon = 1.5$ centered around the v-GNE for agent 4.}
    \label{fig: convergence_time}
\end{figure}

\begin{figure}
    \centering
    \includegraphics[width = \linewidth]{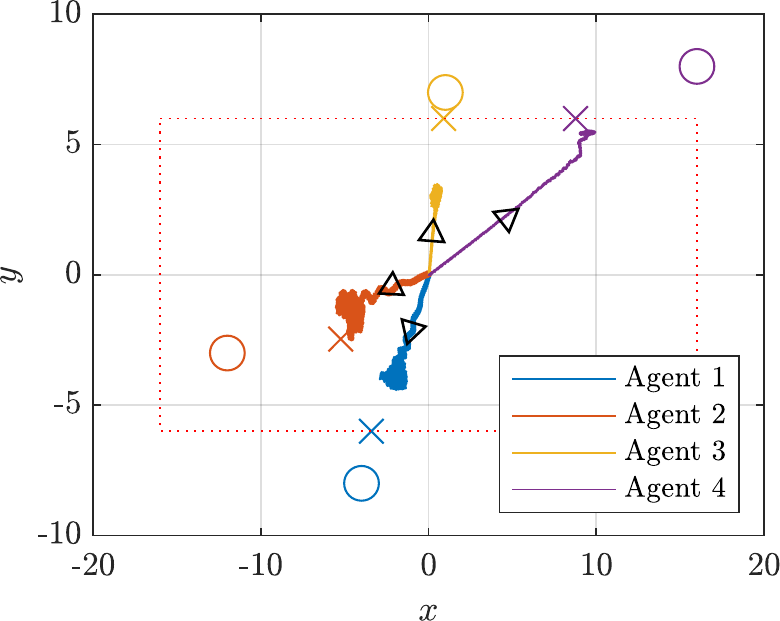}
    \caption{State trajectories in the $x-y$ plane for the case of $d_i = 0.49$. Circle symbols represent locations of the sources, while the $\times$ symbols represent locations of the v-GNE.}
    \label{fig: di = 0.5}
\end{figure}

For our numerical simulations, we choose the parameters as in Table \ref{tab: sim1}. We randomly choose for each of the agents the following perturbation frequencies: $(\overline\omega_1^1, \overline\omega_1^2) = (5.11, 6.38)$, $(\overline\omega_2^1, \overline\omega_2^2) = (4.42, 5.16)$, $(\overline\omega_3^1, \overline\omega_3^2) = (10.59, 11.91)$, $(\overline\omega_4^1, \overline\omega_4^2) = (14.65, 16.12)$.  We run simulations for different values of perturbation amplitudes in range $[0.1, 0.5]$ and different values of the frequency factor $k_\omega$ in range $[0.17, 1]$. The numerical results are illustrated on Figures \ref{fig: convergence_distance}, \ref{fig: convergence_time} and \ref{fig: di = 0.5}. In Figure \ref{fig: convergence_distance}, we see that smaller perturbations and frequency factors bring the system closer towards the v-GNE; however in Figure \ref{fig: convergence_time}, we see that the convergence rate slows down significantly. Thus, there is a trade-off between convergence speed and closeness to the solution. In Figure \ref{fig: di = 0.5}, we see a representative example of agent trajectories for $\|d_i\| = 0.49$. We note that the trajectories of agents 1 and 3 have saturate in their decision variables, while agents 2 and 4 are limited mostly by the coupling constraints.

\begin{table}
\centering
\begin{tabular}{ccccc}
\hhline{=====}
\multicolumn{5}{c}{Problem setup}                                       \\ \hhline{=====}
N          & $x_{\textup{min}}$      & $x_{\textup{max}}$    & $y_{\textup{min}}$       & $y_{\textup{max}}$        \\
4          & -16          & 16         & -6            & 6              \\ \hline
$r_1^s$    & $r_2^s$      & $r_3^s$    & $r_4^s$       & $(c, b)$       \\
$(-4, -8)$ & $(-12, -3)$  & $(1, 7)$   & $(16, 8)$     & $(0.04, 14)$   \\ \hline
\multicolumn{5}{c}{Parameter   estimation scheme}                       \\ \hhline{=====}
$K_i$      & $k_i^\sigma$ & $\sigma_i$ & $\Sigma_i(0)$ & $\omega_i^{j}$ \\
100        & 100          & $10^{-6}$  & $0.1 I$         & $k_\omega\overline\omega_i^j$ \\ \hhline{=====}
\multicolumn{5}{c}{System stabilization   and extremum seeking}         \\ \hhline{=====}
$\gamma_i$ & $\gamma_0$      & $\epsilon$ & $K_1^i$       & $K_2^i$        \\
0.002      & 0.002        & 0.1        & 3             & 6              \\ \hhline{=====}
\end{tabular}
\caption{Simulation parameters for connectivity control.}
\label{tab: sim1}
\end{table}

\subsection{Wind farm optimization}
As one of the main sources of renewable energy, wind farms and their optimization has been addressed extensively from different perspectives such as power tracking of single turbines \cite{koutroulis2006design}, \cite{boukhezzar2006nonlinear}, power tracking via extremum seeking \cite{ghaffari2014extremum}, power tracking with load reduction \cite{soltani2011load}, \cite{soliman2011multiple}, distributed optimization of wind farms  \cite{menon2013distributed}, \cite{marden2013model}, \cite{barreiro2015model} and distributed optimization via extremum seeking \cite{ebegbulem2017distributed}. While in the power tracking case, often the torque or some other related variable is taken as the control input, in the distributed optimization case, the axial induction factor (AIL) is usually taken as the control input. \\ \\
In what follows, we consider a similar problem in which a wind farm tries to maximize its power output with AIL as the control input. The control variables are subject to local constraints (feasible values of AIL). Also, we require that the turbines experience a similar amount of mechanical stress. In order to do that, we impose that AILs of a row of wind turbines cannot differ to much from AILs of the succeeding row, which introduces coupling constraints to the optimization problem. Unlike the previously mentioned literature on distributed wind farm optimization, here we also allow for AIL dynamics in order to reflect the turbine time constant and its effect on the power output. One possible way of solving the problem would be via global optimization, where a central coordinator would minimize a global cost function and send AIL commands to the turbines. To avoid having a single critical node for computation and communication, an alternative approach is to pose the problem as a potential game, where the cost function of the turbines are “aligned" to a global utility function. In our case, the potential function would be the sum of all power outputs. We choose that the individual cost functions are equal to the potential function and each of the agents minimizes their cost function on their own, with limited information from the central coordinator. In this setup, a v-GNE corresponds to an optimal solution of the global power maximization problem.\\ \\
%In \cite{goit2015optimal}, \cite{goit2016optimal}, it was shown that by including the dynamic interaction of the control with the atmospheric boundary layers in the optimization process, it is possible to increase power production. Therefore, it is possible that real by incorporating dynamics into AIL behaviour that the real-life applications of the algorithm would be improved.\\ \\

Technically speaking, we consider $N$ wind turbines, indexed by $i \in \{1, \dots, N \}$, each with the following AIL dynamics and power output:
\begin{align}
    \dot{a}_{i} &= - \tfrac{1}{\tau}({a}_{i} - u_{i}) \\
    y_i &= -\sum_{i \in \mathcal{I}}P_{i}(\bfs{a}) =-\tfrac{1}{2} \rho A \sum_{i \in \mathcal{I}} C_{P}\left(a_{i}\right) V_{i}(\bfs{a})^{3},
\end{align}
where $a_{i}$ represent the state variable, $u_{i}$ represent the control input, namely the AIL reference, $y_i$ is the measured power output of the wind farm, which is broadcasted by the central coordinator, $\rho$ is the air density, $A$ is the surface area encompassed by the blades of a single turbine, $C_{P}\left(a_{i}\right) \coloneqq a_{i}(1 - a_{i})^2$ is the power efficiency coefficient and $V_{i}$ is the average wind speed experienced by wind turbine $i$, as in \cite[Equ. 5]{marden2013model}:
\begin{align}
    V_{i}(a)=U_{\infty}\z{1-2 \sqrt{\sum\left(a_{j} c_{j i}\right)^{2}}}.
\end{align}

The wind turbines are placed in $R$ rows and $C$ columns with coordinates $x_i$ and their indices can be written as $i = i_c + i_r  C$, where $i_c \in \{1, \dots, C\}$ and $i_r \in \{0, \dots, R - 1\}$. They are tasked to maximize the wind farm power output under local constraints $a_i \in [a_{\min}, a_{\max}]$ and coupling constraints $|a_i - a_j| \leq b$ for all $i, j$, where it holds that $j_r = i_r + 1$. \\

\begin{figure}
    \centering
    \def\svgwidth{0.9\linewidth}
%% Accompanies image file 'vjetrenjace.pdf' (pdf, eps, ps)
\begingroup%
  \makeatletter%
  \providecommand\color[2][]{%
    \errmessage{(Inkscape) Color is used for the text in Inkscape, but the package 'color.sty' is not loaded}%
    \renewcommand\color[2][]{}%
  }%
  \providecommand\transparent[1]{%
    \errmessage{(Inkscape) Transparency is used (non-zero) for the text in Inkscape, but the package 'transparent.sty' is not loaded}%
    \renewcommand\transparent[1]{}%
  }%
  \providecommand\rotatebox[2]{#2}%
  \newcommand*\fsize{\dimexpr\f@size pt\relax}%
  \newcommand*\lineheight[1]{\fontsize{\fsize}{#1\fsize}\selectfont}%
  \ifx\svgwidth\undefined%
    \setlength{\unitlength}{310.18852895bp}%
    \ifx\svgscale\undefined%
      \relax%
    \else%
      \setlength{\unitlength}{\unitlength * \real{\svgscale}}%
    \fi%
  \else%
    \setlength{\unitlength}{\svgwidth}%
  \fi%
  \global\let\svgwidth\undefined%
  \global\let\svgscale\undefined%
  \makeatother%
  \begin{picture}(1,0.8898385)%
    \lineheight{1}%
    \setlength\tabcolsep{0pt}%
    \put(0,0){\includegraphics[width=\unitlength,page=1]{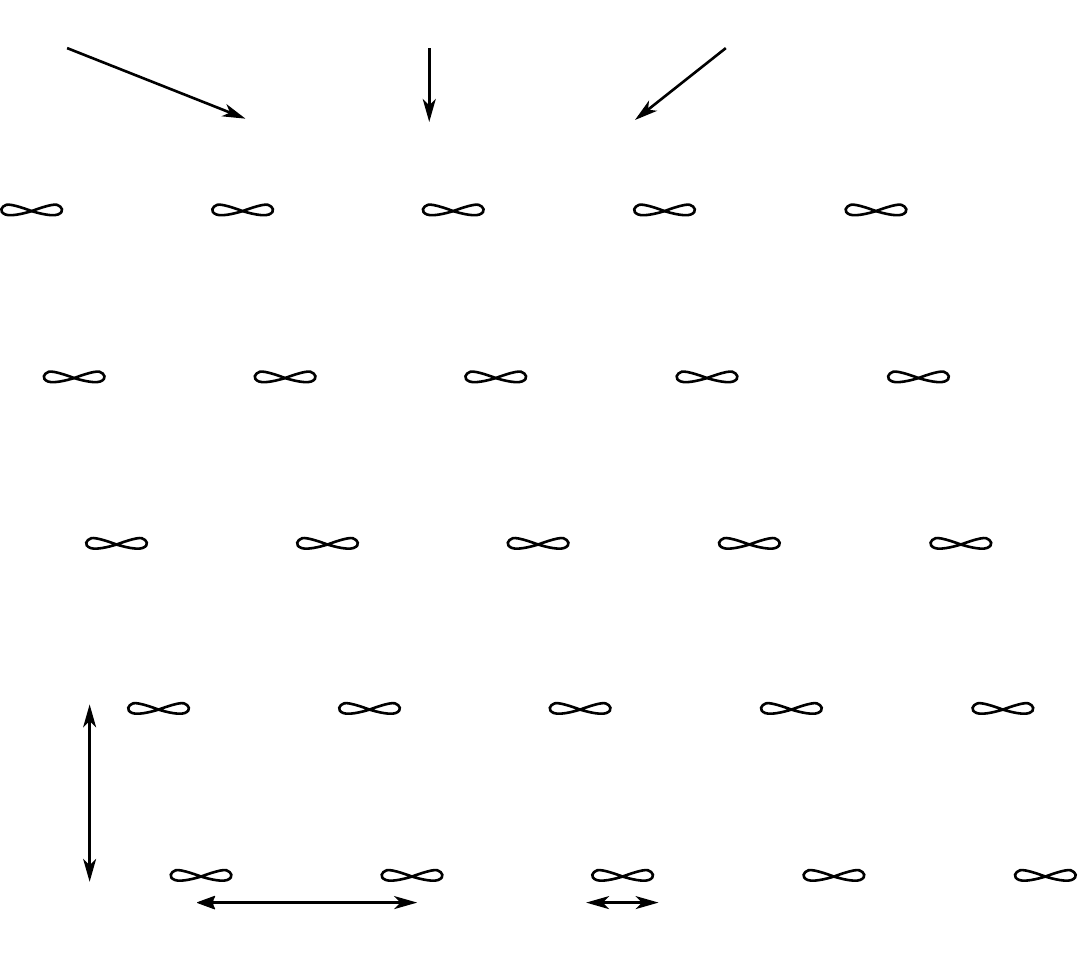}}%
    \put(0.09762056,0.13123962){\color[rgb]{0,0,0}\makebox(0,0)[lt]{\lineheight{1.25}\smash{\begin{tabular}[t]{l}560m\end{tabular}}}}%
    \put(0.24940182,0.00160563){\color[rgb]{0,0,0}\makebox(0,0)[lt]{\lineheight{1.25}\smash{\begin{tabular}[t]{l}560 m\end{tabular}}}}%
    \put(0.55206602,0.00160563){\color[rgb]{0,0,0}\makebox(0,0)[lt]{\lineheight{1.25}\smash{\begin{tabular}[t]{l}80 m\end{tabular}}}}%
    \put(0.07904985,0.85969051){\color[rgb]{0,0,0}\makebox(0,0)[lt]{\lineheight{1.25}\smash{\begin{tabular}[t]{l}$\vec{v}_1$\end{tabular}}}}%
    \put(0.3743621,0.85969051){\color[rgb]{0,0,0}\makebox(0,0)[lt]{\lineheight{1.25}\smash{\begin{tabular}[t]{l}$\vec{v}_2$\end{tabular}}}}%
    \put(0.60941538,0.85969051){\color[rgb]{0,0,0}\makebox(0,0)[lt]{\lineheight{1.25}\smash{\begin{tabular}[t]{l}$\vec{v}_3$\end{tabular}}}}%
    \put(0,0){\includegraphics[width=\unitlength,page=2]{vjetrenjace.pdf}}%
  \end{picture}%
\endgroup%

    \caption{Layout of the wind turbines and the wind directions.}
    \label{fig:windfarm}
\end{figure}

\begin{figure}
    \centering
    \includegraphics{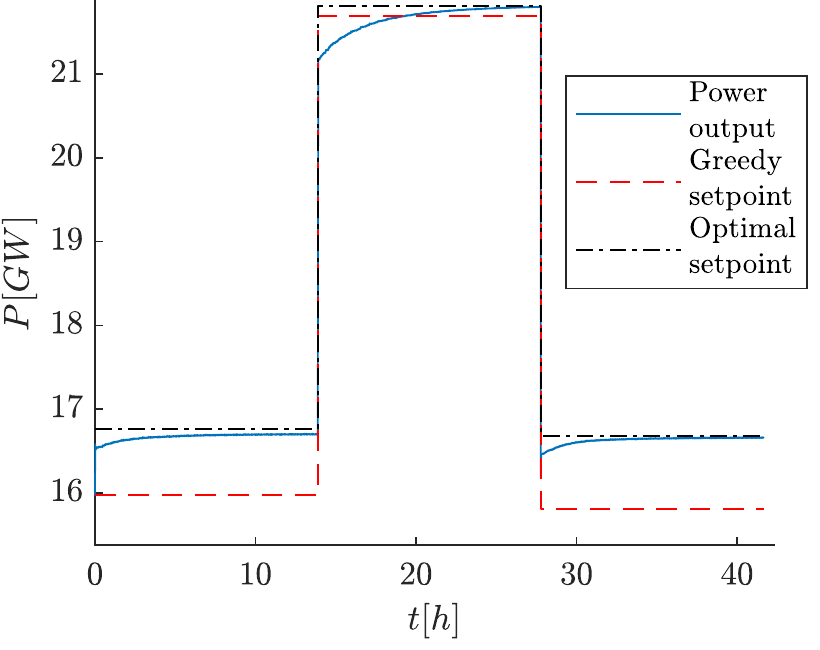}
    \caption{Power generation with the proposed algorithm (solid line) compared to the greedy power output setpoint (dashed red) and the global optimal power setpoint (dot-dashed black).}
    \label{fig: wind}
\end{figure}

For our numerical simulation, we choose a similar setup as in \cite{marden2013model}. The wind farm setup geometric setup is shown in Figure \ref{fig:windfarm} and the following parameters are chosen: $\rho = 1.225$, $U_\infty = 8$, $\tau = 10$, $\gamma_i = \gamma_0 = 0.05$, $\epsilon = 0.005$, $b = 0.03$, $a_{\min} = 0.1$, $a_{\max} = \tfrac{1}{3}$. We take the same parameter estimation scheme as in previous numerical simulation, apart for the perturbation frequencies that we randomly choose in the interval $[3, 11]$ and perturbation amplitudes that we take as $\n{d_i} = 0.01$. All initial conditions, apart for $a_i$, were set to zero. The initial condition for $a_i$ was set to $\tfrac{1}{3}$, which corresponds to the greedy strategy in \cite{marden2013model}. In our simulations, we use three different wind directions. In the time interval $[0, 50000)$, the wind was blowing with speed direction vector $\Vec{v}_1 = (2, -1)$; in the time interval $[50000, 100000)$, the wind was blowing with the speed direction vector $\vec{v}_2 = (0, -1)$; and finally, in the time interval $[100000, 150000]$, the wind was blowing with speed direction vector $\vec{v}_3 = (-1, -1)$. We assume that the wind turbines instantly adjust their orientation towards the wind direction as this process is relatively fast compared with the GNE learning process. The simulation results are shown on Figure \ref{fig: wind}. We can see that the wind turbines reach a neighborhood of the v-GNE, even with the delay introduced by AIL dynamics.

\section{Conclusion}
Generalized Nash equilibrium problems with nonlinear dynamical agents can be solved via a preconditioned forward-backward algorithm that uses estimates of the pseudo-gradient mapping if it is strongly monotone and Lipschitz continuous, and if the dynamical agents have a certain exponential stability property. Regular projections enable the use of a parameter estimation scheme. Numerical simulations show that there is trade-off between closeness to the equilibrium solution and the speed of convergence.
\bibliographystyle{plain}        % Include this if you use bibtex 
%\bibliography{ieeetr}
\bibliography{autosam}           % and a bib file to produce the 
                                 % bibliography (preferred). The
                                 % correct style is generated by
                                 % Elsevier at the time of printing.

%\begin{thebibliography}{99}     % Otherwise use the  
                                 % thebibliography environment.
                                 % Insert the full references here.
                                 % See a recent issue of Automatica 
                                 % for the style.
%  \bibitem[Heritage, 1992]{Heritage:92}
%     (1992) {\it The American Heritage. 
%     Dictionary of the American Language.}
%     Houghton Mifflin Company.
%  \bibitem[Able, 1956]{Abl:56}
%     B.~C.~Able (1956). Nucleic acid content of macroscope. 
%     {\it Nature 2}, 7--9. 
%  \bibitem[Able {\em et al.}, 1954]{AbTaRu:54}   
%     B.~C. Able, R.~A. Tagg, and M.~Rush (1954).
%     Enzyme-catalyzed cellular transanimations.
%     In A.~F.~Round, editor, 
%     {\it Advances in Enzymology Vol. 2} (125--247). 
%     New York, Academic Press.
%  \bibitem[R.~Keohane, 1958]{Keo:58}
%     R.~Keohane (1958).
%     {\it Power and Interdependence: 
%     World Politics in Transition.}
%     Boston, Little, Brown \& Co.
%  \bibitem[Powers, 1985]{Pow:85}
%     T.~Powers (1985).
%     Is there a way out?
%     {\it Harpers, June 1985}, 35--47.

%\end{thebibliography}

\appendix
\section{Proof of Theorem 1}    
To prove the convergence of the algorithm, we show that equation in \nref{eq: control law} is equivalent to a continuous-time preconditioned forward-backward algorithm, whose convergence is proven using well-known properties of monotone operators. First, we show the equivalence. Let us denote $\omega = \operatorname{col}(\bfs{u}, \lambda)$. We write Equation \nref{eq: control law} as: 
\begin{align}
    &\dot{\omega} = -\omega + \operatorname{proj}_{\bfs{\Omega} \times \geq \boldsymbol{0}}\left(\omega + \bfs{\Gamma}\begin{bmatrix} -F(\bfs{u}) - A^\top \lambda \\ A\bfs{u} - b + 2A\dot{\bfs{u}}\end{bmatrix}{} \right), \label{eq: full_info_control_imperfect}
\end{align}
where $\bfs{\Gamma} = \diag{\Gamma, \gamma_0 I_q}$. Next, by the property of projection operator in \cite[Prep. 6.47]{bauschke2011convex}, Equation \nref{eq: full_info_control_imperfect} reads as
\begin{align}
    &\dot{\omega} + \omega = (\operatorname{Id} + \bfs{\Gamma}\text{N}_{\bfs{\Omega} \times \geq \boldsymbol{0}})^{-1}\left(\omega + \bfs{\Gamma}\begin{bmatrix} -F(\bfs{u}) - A^\top \lambda \\ A\bfs{u} - b + 2A\dot{\bfs{u}}\end{bmatrix}{} \right), \nonumber
\end{align}
which is equivalent to the inclusion 
\begin{align}
    &\dot{\omega} + \omega + \bfs{\Gamma}\text{N}_{\bfs{\Omega} \times \geq \boldsymbol{0}}(\dot{\omega} + \omega) \ni \omega + \bfs{\Gamma}\begin{bmatrix} -F(\bfs{u}) - A^\top \lambda \\ A\bfs{u} - b + 2A\dot{\bfs{u}}\end{bmatrix}{}.\label{eq: full_info_control_semiperfect}
\end{align}
When the elements of the last matrix product in \nref{eq: full_info_control_semiperfect} are rearranged and the equation is premultiplied by $\bfs{\Gamma}^{-1}$, the equations read as follows:
\begin{align}
    &\dot{\omega} + \omega + \bfs{\Gamma}\text{N}_{\bfs{\Omega} \times \geq \boldsymbol{0}}(\dot{\omega} + \omega) \ni \tilde{\Gamma}\omega + \begin{bmatrix}-\Gamma F(\bfs{u}) \\ -\gamma_0 b\end{bmatrix} + \begin{bmatrix} 0 \\ \gamma_0 2A\dot{\bfs{u}} \end{bmatrix} \Leftrightarrow \nonumber\\
    &\bfs{\Gamma}^{-1}(\dot{\omega} + \omega) + \text{N}_{\bfs{\Omega} \times \geq \boldsymbol{0}}(\dot{\omega} + \omega) \ni \hat{\Phi}\omega + \begin{bmatrix}- F(\bfs{u}) \\ - b\end{bmatrix} + \begin{bmatrix} 0 \\  2A\dot{\bfs{u}} \end{bmatrix}, \label{eq: full_info_control_perfect}
\end{align}
where we have used the notation $\tilde{\Gamma} = \begin{bmatrix} I_m & -\Gamma A^\top \\ \gamma_0 A & I_q \end{bmatrix}$ and  $\hat{\Phi} = \begin{bmatrix} \Gamma^{-1} &  -A^\top \\  A & \gamma_0^{-1}I_q \end{bmatrix}$. Now, since
\begin{align}
    &\begin{bmatrix} \Gamma^{-1} &  -A^\top \\  A & \gamma_0^{-1}I_q \end{bmatrix}\omega  + \begin{bmatrix} 0 \\  2A\dot{\bfs{u}} \end{bmatrix}\Leftrightarrow \nonumber\\
    & \begin{bmatrix} \Gamma^{-1} &  -A^\top \\  -A & \gamma_0^{-1}I_q \end{bmatrix}\omega + \begin{bmatrix} 0 &  0 \\  2A & 0 \end{bmatrix}\omega + \begin{bmatrix} 0 & 0\\ 2 A & 0\end{bmatrix} \dot{\omega} \Leftrightarrow \nonumber\\
    & \begin{bmatrix} \Gamma^{-1} &  -A^\top \\  -A & \gamma_0^{-1}I_q \end{bmatrix}\omega + \begin{bmatrix} 0 & 0 \\ 2 A & 0 \end{bmatrix}(\omega + \dot{\omega})\Leftrightarrow \nonumber\\
        & \Phi\omega + \Hat{A}(\omega + \dot{\omega}),
\end{align}{}
%where ${\Phi} = \begin{bmatrix} \gamma^{-1} &  -A^\top \\  -A & \nu^{-1} \end{bmatrix}$ and $\Hat{A} = \begin{bmatrix} 0 & 0 \\ 2 A & 0 \end{bmatrix}$. 
Equation \nref{eq: full_info_control_perfect} reads as
\begin{align}
    &\bfs{\Gamma}^{-1}(\dot{\omega} + \omega) + \text{N}_{\bfs{\Omega} \times \geq \boldsymbol{0}}(\dot{\omega} + \omega) \ni \nonumber \\
    &\Phi\omega + \begin{bmatrix}- F(\bfs{u}) \\ - b\end{bmatrix} + \Hat{A}(\omega + \dot{\omega}). \label{eq: full_info_control_superperfect}
\end{align}Next, the following expression is valid for the matrices:
\begin{align}
    \bfs{\Gamma}^{-1} - \hat{A} = \Phi + \m{0 & A^\top \\ -A & 0} = \Phi + \Psi. \label{eq: matrix miracle}
\end{align}{}
From Equations \nref{eq: full_info_control_superperfect} and \nref{eq: matrix miracle}, it follows
\begin{align}
    \Phi(\dot{\omega} + \omega)  + \Psi (\omega + \dot{\omega}) + \text{N}_{\bfs{\Omega} \times \geq \boldsymbol{0}}(\dot{\omega} + \omega) \ni 
    \Phi\omega - \begin{bmatrix}F(\bfs{u}) \\ b\end{bmatrix} \Leftrightarrow \nonumber\\
    (\dot{\omega} + \omega) + \Phi^{-1}\left(\text{N}_{\bfs{\Omega} \times \geq \boldsymbol{0}} + \Psi \right)(\dot{\omega} + \omega)\ni \omega - \Phi^{-1}\begin{bmatrix}F(\bfs{u}) \\ b\end{bmatrix}.   
\end{align}
By inverting the operator on the left side of the previous expression, we finally arrive to desired equation:
\begin{align}
    &\dot{\omega} = - \omega \nonumber \\
    &+ \left(\operatorname{Id} + \Phi^{-1} \left(\text{N}_{\bfs{\Omega} \times \geq \boldsymbol{0}} + \Psi\right)\right)^{-1} \left(\omega - \Phi^{-1}\begin{bmatrix}F(\bfs{u}) \\ b\end{bmatrix} \right) \nonumber \Leftrightarrow \nonumber\\
    &\dot{\omega} = - \omega + \left(\operatorname{Id} + \Phi^{-1} \mathcal{A}\right)^{-1} \circ \left(\omega - \Phi^{-1}\mathcal{B}(\omega) \right) \Leftrightarrow \nonumber\\
    &\dot{\omega} = - \omega + J_{\Phi^{-1} \mathcal{A}}\left(\omega - \Phi^{-1}\mathcal{B}(\omega) \right). \label{last equation}
\end{align}{}
Equation \nref{last equation} represents a forward-backward algorithm preconditioned by matrix $\Phi^{-1}$. Fixed points of the operator on the right-hand side of \nref{last equation} (\cite[Prep. 26.1, (iv)(a)]{bauschke2011convex}, \cite[Lemma 1]{yi2019operator}, \nref{equ: kkt1}, \nref{equ: kkt2}) represent Generalized Nash equilibria that are the solutions to the game in \nref{def: dyn_game}. Before proving convergence, we have to prove an additional result.\\

\begin{lem}{} \label{lemma: jedina} Let $T = (\operatorname{Id} + \mathcal{A})^{-1}\circ(\operatorname{Id} - \mathcal{B})$, where $\mathcal{A}$ is maximally monotone. Then it holds:
\begin{align}
    (Tx - x^*)^\top (x - Tx ) \geq (Tx - x^*)^\top (\mathcal{B}x - \mathcal{B}x^*).\nonumber
\end{align}{}
for all $(x, x^*) \in \operatorname{dom}(T)\times \operatorname{fix}(T)$.\kraj
\end{lem}
\begin{pf}
Let us denote $x^* = Tx^* = J_\mathcal{A} y^*$ as the fixed point of operator $T$. Then it holds:
\begin{align}
    &\ (Tx - x^*)^\top (x - Tx - (\mathcal{B}x - \mathcal{B}x^*))\nonumber \\
    =&\ (Tx - x^*)^\top (x - \mathcal{B}x -  Tx + x^* - (x^* - \mathcal{B}x^*))\nonumber \\
    =&\ (Tx - x^*)^\top (y - Tx + x^* - y^*)\nonumber \\
    =&\ (J_\mathcal{A} y - J_\mathcal{A} y^*)^\top ((\operatorname{Id} - J_\mathcal{A})y - (\operatorname{Id} - J_\mathcal{A})y^*)\nonumber \\
    \geq &\ 0,
\end{align}{}where the last equation holds due to properties of firmly nonexpansive operators. \hfill $\blacksquare$\end{pf}
Now we denote $\Tilde{\mathcal{A}} = \Phi^{-1} \mathcal{A}$, $\Tilde{\mathcal{B}} = \Phi^{-1} \mathcal{B}$ and $T = (\operatorname{Id} + \Tilde{\mathcal{A}})^{-1}\circ(\operatorname{Id} - \Tilde{\mathcal{B}})$. Then, the dynamics in \nref{last equation} read as
\begin{align}
    \dot{\omega} = -\omega + T\omega. \ \label{eq: continous_dynamics}
\end{align}{}
We propose the Lyapunov function candidate
\begin{align}
V(\omega) = \dfrac{1}{2}\| \omega - \omega^*\|^2,
\end{align}{}
where $\omega^*$ is a fixed point of operator $T$.\ Its derivative along the trajectory in \nref{last equation} is then
\begin{align}
    \dot{V}(\omega) &= -(\omega - \omega^*)^\top (\omega - T\omega) \nonumber\\
    &= -\| (\omega - T\omega)\|^2 -(T\omega - \omega^*)^\top (\omega - T\omega)\nonumber\\
    &= -\| \dot{\omega}\|^2 -(T\omega - \omega^*)^\top (\omega - T\omega)
\end{align}{}
From Lemma \ref{lemma: jedina}, it follows that
\begin{align}
        \dot{V}(\omega) &\leq -\| \dot{\omega}\|^2 -(T\omega - \omega^*)^\top (\Tilde{\mathcal{B}}\omega - \Tilde{\mathcal{B}}\omega^*)\nonumber \\
        & \leqq -\|\dot{\omega}\|^2 -(T\omega - \omega)^\top (\Tilde{\mathcal{B}}\omega - \Tilde{\mathcal{B}}\omega^*) \nonumber\\ 
        &\quad  - (\omega - \omega^*)^\top (\Tilde{\mathcal{B}}\omega - \Tilde{\mathcal{B}}\omega^*)\nonumber \\
        & \leqq -\| \dot{\omega}\|^2 -(T\omega - \omega)^\top \Phi^{-1}(\mathcal{B}\omega - \mathcal{B}\omega^*) \nonumber\\
        &\quad - (\omega - \omega^*)^\top \Phi^{-1}(\mathcal{B}\omega - \mathcal{B}\omega^*). \label{eq: vdot_imperfect}
\end{align}{}
Bounds on the eigenvalues of $\Phi$ can be estimated with Gershgorin’s theorem. For small enough step sizes, we denote the lower and upper bound on the eigenvalues as $\sigma_{\textup{min}} = \frac{1}{\max_{i \in \mathcal{I}_0}(\gamma_i^{-1}) + \|A\|}$ and $\sigma_{\textup{max}} = \frac{1}{\min_{i \in \mathcal{I}_0}(\gamma_i^{-1}) - \|A\|}$, respectively. We bound \nref{eq: vdot_imperfect} as
\begin{align}
  \dot{V}(\omega) \leq &-\| \dot{\omega}\|^2 + \sigma_{\textup{max}}\|\dot{\omega}\| \|\mathcal{B}\omega - \mathcal{B}\omega^*\|  \nonumber \\
  & -\sigma_{\textup{min}}(\omega - \omega^*)^\top (\mathcal{B}\omega - \mathcal{B}\omega^*). \label{eq: vdot_semiperfect}  
\end{align}{}
Since $F(\bfs{u})$ is strongly monotone and Lipschitz continuous, it is also cocoercive \cite[Lemma 5]{yi2019operator}. Therefore, the operator $\mathcal{B}$ is cocoercive with constant $\beta = \frac{\mu}{L^2}$. Equation \nref{eq: vdot_semiperfect} then becomes:
\begin{align}
  &\dot{V}(\omega) \leq -\| \dot{\omega}\|^2 + \sigma_{\textup{max}}\|\dot{\omega}\| \|\mathcal{B}\omega - \mathcal{B}\omega^*\| \nonumber \\
  &- \frac{\sigma_{\textup{min}}}{2}\beta \|\mathcal{B}\omega - \mathcal{B}\omega^*\|^2 - \frac{\sigma_{\textup{min}}}{2}(\omega - \omega^*)^\top (\mathcal{B}\omega - \mathcal{B}\omega^*)\nonumber\\
  &\leq -\frac{1}{2}\| \dot{\omega}\|^2 - \frac{\sigma_{\textup{min}}}{2}(\omega - \omega^*)^\top (\mathcal{B}\omega - \mathcal{B}\omega^*) \nonumber \\
  &- \frac{1}{2} \begin{bmatrix} \| \dot{\omega}\| & \|\mathcal{B}\omega - \mathcal{B}\omega^*\|\end{bmatrix} \begin{bmatrix} 1 & \sigma_{\textup{max}} \\ \sigma_{\textup{max}} & \beta \sigma_{\textup{min}}\end{bmatrix}  \begin{bmatrix}\|\dot{\omega}\| \\ \|\mathcal{B}\omega - \mathcal{B}\omega^*\| \end{bmatrix} \label{eq: vdot_perfect}
\end{align}{}
Since, it is always possible to choose parameters $\gamma_i$ and $\gamma_0$ small enough such that $\beta \sigma_{\textup{min}} \geq \sigma_{\textup{max}}^2$ and the matrix in \nref{eq: vdot_perfect} is negative definite, the last equation reads as 
\begin{align}
  \dot{V}(\omega) &\leq -\frac{1}{2}\| \dot{\omega}\|^2 - \frac{\sigma_{\textup{min}}}{2}(\omega - \omega^*)^\top (\mathcal{B}\omega - \mathcal{B}\omega^*)\nonumber \\
  &\leq -\frac{1}{2}\| \dot{\omega}\|^2 - \frac{ \mu \sigma_{\textup{min}}}{2}\|\tilde{\bfs{u}}\|^2,
\end{align}{}
where $\tilde{\bfs{u}} = \bfs{u} - \bfs{u}^*$ and the last line follows from strong monotonicity of $F(\bfs{u}^*)$. Rest of the proof represent a La Salle argument. As the right-hand side is a sum of negative squares, it follows that $\dot{V}(\omega) \leq 0$ for all $\omega$. Let us denote $\zeta_0 = \{\omega\ \in \mathbb{R}^{n + q}\ |\ \dot{V}(\omega) = 0 \}$. Then, it holds $\zeta_0 \subseteq \zeta_1 = \{\omega\ \in \mathbb{R}^{n + q}\ |\ \|\dot{\omega}\| = 0\text{ and }\|\tilde{\bfs{u}}\| = 0\}$. When $\dot{\omega}\ = 0$, from the dynamics \nref{eq: continous_dynamics}, it follows that
\begin{align}
    \omega = T(\omega).
\end{align}{}i.e., $\omega\ \in \operatorname{fix}(T)$. The set $\zeta_1$ can be rewritten as $\zeta_1 = \{(\bfs{u}, \lambda)\ \in \operatorname{fix}(T)\ | \ \bfs{u} = \bfs{u}^*\}$. As $\omega^*$ was chosen as an arbitrary fixed point of the operator $T$, it follows that $u^*$ is a singleton and $\zeta_1 = \operatorname{fix}(T)$. From the dynamics \nref{eq: continous_dynamics}, it follows that the largest invariant set is $\zeta_{\mathrm{inv}}  = \zeta_1 =  \operatorname{fix}(T)$. Therefore, $\zeta_0 \subseteq \zeta_{\mathrm{inv}}$. By La Salle's invariance principle \cite[Thm. 4.4]{khalil2002nonlinear}, we conclude that the state trajectories converge to the set $\zeta_0$, in which $u = u^*$ is a singleton and it holds that $\operatorname{col}(u^*, \lambda) \in \operatorname{fix}(T)$.\\ \\

\section{Proof of Theorem 2}
Let us consider a Lyapunov function candidate of the form $V = V_\theta + V_\omega$, where $V_\theta$ represents a parameter estimation error and $V_\omega$ represents a primal-dual convergence error:
\begin{align}
   & V_\theta(\tilde{\bfs{\eta}}, \tilde{\bfs{\theta}}) =\textstyle\sum_{i \in \mathcal{I}}\left(\frac{1}{2} \widetilde{\eta}_{i}^\top \widetilde{\eta}_{i}+\frac{1}{2} \widetilde{\theta}_{i}^\top \Sigma_{i} \widetilde{\theta}_{i}\right), \label{eq: lyapunov parameter est}\\
    &V_\omega(\omega) = \frac{1}{2}\|\omega - \omega^*\|^2\label{eq: V_i}.
\end{align}{}
Since we anticipate that the derivative of the projection function does not exist on some corner points, we use the Lyapunov theory for differential inclusions as in \cite[Chp. 2]{blanchini2008set}, namely we use upper Dini derivatives ($D^+$) instead of regular time derivatives. For ease of notation, we use the regular derivatives whenever they are equal to Dini derivatives.\\ \\
\emph{Outline of the proof:} We first bound all of the positive terms in $D^+{V_\theta}$ with functions of variables $(\bfs{\eta}, \hat{\bfs{\theta}}, \omega)$, then we similarly bound all of the terms in $D^+{V_\omega}$ introduced by the parameter estimates. Finally, we use the quadratic terms of $D^+ V$ to show that the positive terms are majorized by the negative terms.\\ \\

\emph{Parameter estimation term:} We bound the Dini derivative of $V_\theta$ similarly to \cite[Thm. 1]{guay2017proportional} and \cite[Eq. 31]{guay2018distributed} with the only difference that we let each agent choose their own parameters ($\sigma_i, K_i, \rho_i$). The Lyapunov derivative reads as follows:
\begin{align}
    &D^+{V_\theta}(\tilde{\bfs{\eta}}, \tilde{\bfs{\theta}}) \leq \sum_{i \in \mathcal{I}}\Bigg (-\widetilde{\eta}_{i}^\top\left(K_i-\frac{1}{2}-\frac{k_{1} \zeta_i}{2}\right) \widetilde{\eta}_{i} \nonumber \\ 
    & - \frac{1}{2}\|e_i - \eta_i\|^2 + \frac{k_{i}^{\Sigma\prime} \gamma_{1i}}{2} \widetilde{\theta}_{i}^{\tau} \widetilde{\theta}_{i}+\frac{\sigma_i}{2} \theta_{i}^\top \theta_{i}\nonumber \\
    & +\frac{1}{2 k_{1}} D^+ \theta_i^\top D^+ \theta_i  +\frac{\gamma_{2i}}{2 k_{2}} D^+ \theta_i^\top D^+ \theta_i  \Bigg )\nonumber \\
    \leq& - k_\textup{a} \|\tilde{\bfs{\eta}} \|^2 - k_\textup{b}\| \tilde{\bfs{\theta}}\|^2 - \frac{1}{2}\|\bfs{e} - \bfs{\eta}\|^2+ k_\textup{c} \| D^+{\bfs{\theta}} \|^2\nonumber \\
    &+ \frac{\sigma}{2}\|\bfs{\theta}\|^2, \label{eq: bound W}
\end{align}where $k_\textup{a} \coloneqq \min_i \left(K_i-\frac{1}{2}-\frac{k_{1} \zeta_i}{2}\right)$, $k_\textup{b} \coloneqq \min_i \left(\frac{k_{i}^{\Sigma \prime} \gamma_{1i}}{2}\right)$, $k_\textup{c} \coloneqq \max_i \left(\frac{1}{2k_1} + \frac{\gamma_{2i}}{2k_2}\right)$ and $\sigma \coloneqq\max_i\sigma_i$\footnote{The term $\frac{1}{2}\|\bfs{e} - \bfs{\eta}\|^2$ was omitted in \cite{guay2017proportional} (look at page 4, first column, second to last equation) and \cite{guay2018distributed}, as it wasn't required.}.\\ \\
Next, we bound the positive terms in \nref{eq: bound W}. The analysis starts with $\bfs{\theta} = \col{\bfs{\theta}^0, \bfs{\theta}^1}$, where
\begin{align}
    \bfs{\theta}^0&\coloneqq\operatorname{col}\left(\left(\nabla_{\bfs{u}_{-i}} J_{i}\left(u_{i}, \bfs{u}_{-i}\right)^\top \dot{\bfs{u}}_{-i}\right)_{i \in \mathcal{I}}\right) \nonumber\\
    &\coloneqq J^0(\bfs{u})\dot{\bfs{u}}, \label{def: theta_0} \\
    \bfs{\theta}^1 &\coloneqq F(\bfs{u}).
\end{align}We have that
\begin{align}
    \| \bfs{\theta}^0 \| &\leq \|J^0(\bfs{u})\| \|\dot{\bfs{u}}\| = L^0  \|\dot{\bfs{u}}\| , \nonumber \\
    \| \bfs{\theta}^1 \| &= \|F(\bfs{u})\| \leq \|F(\bfs{u}) - F(\bfs{u}^*)\| + \|F(\bfs{u}^*)\| \nonumber \\
    &\leq \ell\|\tilde{\bfs{u}}\| + \|F(\bfs{u}^*)\|, \nonumber
\end{align}
where $L^0 \coloneqq \max_{\bfs{u} \in \mathcal{U}} \n{J^0(\bfs{u})} < \infty$, since $\mathcal{U}$ is bounded. Then, we bound $\|\theta\|$ as follows %in all arguments for $i \in \mathcal{I}$
\begin{align}
    \|\theta\|^2 &\leq {L^0}^2  \|\dot{\bfs{u}}\|^2 + (\ell\|\tilde{\bfs{u}}\| + \|F(\bfs{u}^*)\|)^2 \nonumber \\
    %&\leq 3({L^0}^2  \|\dot{\bfs{u}}\|^2 + L^2\|\tilde{\bfs{u}}\|^2 + \|F(\bfs{u}^*)\|^2)\nonumber \\
    & \leqq L_1\|\dot{\bfs{u}}\|^2 + L_2\|\tilde{\bfs{u}}\|^2 + L_3\|F(\bfs{u}^*)\|^2, \nonumber
\end{align}
for $L_1 \coloneqq {L^0}^2 $, $L_2 \coloneqq 2 \ell^2$ and $ L_3 = 2$. In order to bound $\|D^+{\bfs{\theta}}\|$, we observe the dini derivatives of $\bfs{\theta}^0$ and $\bfs{\theta}^1$:
\begin{align}
    &\|D^+\bfs{\theta}^0\|\leq \|\dot{\bfs{u}}^\top H_{J^0}\dot{\bfs{u}}\| + \| J^0(\bfs{u}) D^+\dot{\bfs{u}}\|,\label{eq: bound theta0 imperfect}\\ 
    &\|D^+\bfs{\theta}^1\| = \|\nabla F(\bfs{u}) \dot{\bfs{u}}\| \leq L \|\dot{\bfs{u}}\|, \label{eq: bound theta1 imperfect}\\
    &\|D^+\dot{\bfs{u}}\| = \n{\dot{\bfs{u}} + D^+ \operatorname{proj}_{\bfs{\Omega}}\left(\bfs{u} - \Gamma (\bfs{\theta}^1 + A^\top\lambda) + d(t) \right)} \nonumber\\
    &\quad\leq \| \dot{\bfs{u}}\| + \n{D^+ \operatorname{proj}_{\bfs{\Omega}}\left( \bfs{u} + \Gamma (\hat{\bfs{\theta}}^1 - A^\top\lambda) + d(t) \right) } \nonumber\\ 
    &\quad\leq 2\|\dot{\bfs{u}}\| + \sigma_{\textup{max}}(\Gamma)\|\dot{\hat{\bfs{\theta}}}^1\| + \sigma_{\textup{max}}(\Gamma)\|A\| \|\dot{\lambda}\| + \|\dot{d}(t)  \| \nonumber
\end{align}{}
% We note that the projection operator need not be differentiable on the boundary of the convex set. For example, $\proj_{\geq 0}(t)$ is not differentiable at $t = 0$ as $\lim_{\delta t \rightarrow 0^-}\frac{\proj_{\geq 0}(t + \delta t) - \proj_{\geq 0}(t)}{\delta t} = 0$ and $\lim_{t \rightarrow 0^+} \frac{\proj_{\geq 0}(t + \delta t) - \proj_{\geq 0}(t)}{\delta t}= 1$. Although the \emph{limit} does not exist, left and right-hand limits \emph{must} exist and they are bounded easily as:
% \begin{align}
%     \lim_{\delta t \rightarrow 0^+, 0^-}\n{\frac{\proj_{\Omega}(f(t + \delta t)) - \proj_{\Omega}(f(t))}{\delta t}}  \nonumber\\  \leq \lim_{\delta t \rightarrow 0}\n{\frac{f(t + \delta t) - f(t)}{\delta t}} = \n{\frac{d}{dt}f(t)}.
% \end{align}
% Therefore the Dini derivative is also bounded. 
Next, we bound $\|\dot{\hat{\bfs{\theta}}}^1\|$ using the estimator dynamics in \nref{estimlast}:
\begin{align}
    \|\dot{\hat{\bfs{\theta}}}\| \leq &\sum_{i \in \mathcal{I}}(\| \Sigma_i^{-1}c_i(e_i - \eta_i)\| + \sigma_i\|\Sigma_i^{-1}\| \|\theta_i \| \nonumber \\
    &+ \sigma_i\|\Sigma_i^{-1}\| \|\tilde{\theta_i} \|) \nonumber
\end{align}{}
On a compact set $\Omega_c \coloneqq \{(\omega, \hat{\bfs{\eta}}, \hat{\bfs{\theta}}) \in \bfs{\mathcal{{U}}}  \times \R^{q} \times \R^N \times \Theta \ |\  V(\omega, \tilde{\bfs{\eta}}, \tilde{\bfs{\theta}}) \leq c\}$, $c_i$ and $\Sigma_i$ are bounded, therefore, the last equation reads as:
\begin{align}
    \|\dot{\hat{\bfs{\theta}}}\| \leq L_3^*\|(\bfs{e} - \bfs{\eta})\| + L_4^* \|\bfs{\theta} \| + L_5^* \|\tilde{\bfs{\theta}} \|, \nonumber
\end{align}{}
for some positive $L_3^*, L_4^*, L_5^*$. Now, bound on $\|D^+\dot{\bfs{u}}\|$ equals to:
\begin{align}
    \|D^+\dot{\bfs{u}}\|\leq 2\|\dot{\bfs{u}}\| + \sigma_{\textup{max}}(\Gamma)L_3^*\|(\bfs{e} - \bfs{\eta})\| + \sigma_{\textup{max}}(\Gamma)L_4^*\| \|\bfs{\theta} \| \nonumber \\ + \sigma_{\textup{max}}(\Gamma)L_5^* \|\tilde{\bfs{\theta}}\| 
    + \sigma_{\textup{max}}(\Gamma)\|A\| \|\dot{\lambda}\| + \|\dot{d}(t)  \| \label{eq: bound ddot u}
\end{align}{}
On a compact set $\Omega_c$, $\dot{\bfs{u}}^\top H_{J^0}$ is bounded, therefore by combining \nref{eq: bound theta0 imperfect}, \nref{eq: bound theta1 imperfect}, \nref{eq: bound ddot u} and the arithmetic mean - quadratic mean inequality, it follows:
\begin{align}
    \| D^+{\bfs{\theta}}\|^2 &\leq L_4 \| \dot{\bfs{u}} \|^2 + L_5 \| \dot{\lambda} \|^2 + L_6 \| \bfs{\theta} \|^2 +  L_7 \| \tilde{\bfs{\theta}} \|^2\nonumber \\
    &+  L_8 \| \bfs{e} - \bfs{\eta} \|^2 +   L_9 \| \dot{d}(t) \|^2, \nonumber
\end{align}{}
for some positive $L_4$ to $L_9$. By using the previously calculated bounds, $D^+{V_\theta}$ reads as:
\begin{align}
    &D^+{V_\theta} \leq -k_{\textup{a}}\|\tilde{\eta}\|^{2}-(k_{\textup{b}} - k_{\textup{c}} L_7)\|\tilde{\bfs{\theta}}\|^{2}  \nonumber \\
    &- \z{\tfrac{1}{2}-k_\textup{c} L_8}\|\bfs{e} - \bfs{\eta}\|^2+ (k_\textup{c} L_4 + (\sigma + k_\textup{c} L_6)L_1) \| \dot{\bfs{u}} \|^2  \nonumber \\
    &+ (\sigma + k_\textup{c} L_6)L_2 \|\tilde{\bfs{u}} \|^2 + k_\textup{c} L_5 \| \dot{\lambda}\|^2 + k_\textup{c} L_9 \| \dot{d}(t) \|^2 \nonumber \\
    &+ (\sigma + k_\textup{c} L_6)L_3 \|F(\bfs{u}^*) \|. \label{eq: dot w semiperfect}
\end{align}{}

\emph{Primal-dual term:} Unlike the full-information case, our agents use the estimate $\hat{\bfs{\theta}}^1$ instead of $F(\bfs{u})$ in the control law in \nref{eq: control law}. Therefore, by adding and subtracting the derivative of full-information case in the Dini derivative of \nref{eq: lyapunov parameter est}, we have:
\begin{align}
    &D^+{V_\omega}(\omega) = (\omega - \omega^*)^\top\Bigg[-\omega + J_{\Phi^{-1} \mathcal{A}}\left(\omega - \Phi^{-1}\mathcal{B}(\omega) \right) \nonumber \\
    &- \operatorname{proj}_{\bfs{\Omega} \times \geq \boldsymbol{0}}\left(\omega + \bfs{\Gamma}\begin{bmatrix} -F(\bfs{u}) - A^\top \lambda \\ A\bfs{u} - b + 2A\dot{\bfs{u}}\end{bmatrix}{} \right) \nonumber\\ 
   & + \operatorname{proj}_{\bfs{\Omega} \times \geq \boldsymbol{0}}\left(\omega + \bfs{\Gamma}\begin{bmatrix} -\bfs{\theta}^1 - A^\top \lambda \\ A\bfs{u} - b + 2A\dot{\bfs{u}}\end{bmatrix} + \begin{bmatrix}d(t)\\ 0\end{bmatrix} \right)\Bigg]\nonumber \\
    &\leq -\frac{1}{2}\| \dot{\omega}\|^2 - \frac{ \mu \sigma_{\textup{min}}}{2}\|\tilde{\bfs{u}}\|^2 + \sigma_{\textup{max}}(\Gamma)\|\tilde{\bfs{u}}\| \|\tilde{\bfs{\theta}}\| + \|\tilde{\bfs{u}}\|\|d(t)\|\nonumber \\
    &\leq -\frac{1}{2}\| \dot{\omega}\|^2 - \z{\frac{ \mu \sigma_{\textup{min}}}{2} - \frac{\sigma_{\textup{max}}(\Gamma)}{2 k_3} - \frac{1}{2 k_4}}\|\tilde{\bfs{u}}\|^2 \nonumber\\
    &+ \frac{\sigma_{\textup{max}}(\Gamma) k_3}{2} \| \tilde{\bfs{\theta}} \|^2 + \frac{k_4}{2} \|d(t)\|^2 ,\nonumber
\end{align}{}
where the last line follows from the inequality
\begin{align}
    &\n{x} \n{y} \leq \dfrac{1}{2k}\n{x}^2 + \dfrac{k}{2}\n{y}^2 & \forall(x, y, k) \in (\R^{2n}\times \R_{>0} ). \label{eq: famous_ineq}
\end{align}
\emph{Complete Lyapunov candidate:} Finally, the Dini derivative of $V$ is bounded as follows:
\begin{align}
    &D^+{V_\omega} + D^+{V_\theta} \leq  -\z{\frac{1}{2} - k_\textup{c} L_4 - (\sigma + k_\textup{c} L_6)L_1}\| \dot{\bfs{u}}\|^2 
    \nonumber \\
    &-\z{\frac{1}{2} - k_\textup{c} L_5}\| \dot{\lambda}\|^2 - k_{\textup{a}}\|\tilde{\eta}\|^{2}- \z{\frac{1}{2}-k_\textup{c} L_8}\|\bfs{e} - \bfs{\bfs{\eta}}\|^2  \nonumber \\
    &- \z{\frac{ \mu \sigma_{\textup{min}}}{2} - \frac{\sigma_{\textup{max}}(\Gamma)}{2 k_3} - \frac{1}{2 k_4} - (\sigma - k_\textup{c} L_6)L_2}\|\tilde{\bfs{u}}\|^2  \nonumber \\
    & - \z{k_{\textup{b}} - k_{\textup{c}} L_7 - \frac{\sigma_{\textup{max}}(\Gamma) k_3}{2}} \|\tilde{\bfs{\theta}}\|^{2} \nonumber \\
    &+ k_\textup{c} L_9 \| \dot{d}(t) \|^2 + (\sigma + k_\textup{c} L_6)L_3 \|F(\bfs{u}^*) \|^2+ \frac{k_4}{2} \|d(t)\|^2\nonumber \\
    & \leqq -b_1 \| \dot{\bfs{u}}\|^2  -b_2 \| \dot{\lambda}\|^2 -  k_{\textup{a}}\|\tilde{\bfs{\eta}}\|^{2} b_3 - b_3 \|\bfs{e} - \bfs{\eta}\|^2  - b_4 \|\tilde{\bfs{u}}\|^2 \nonumber\\
    &  - b_5 \|\tilde{\bfs{\theta}}\|^{2} + k_\textup{c} L_9 \| \dot{d}(t) \|^2 + (\sigma + k_\textup{c} L_6)L_3 \|F(\bfs{u}^*) \|^2\nonumber \\ 
    &+ \frac{k_4}{2} \|d(t)\|^2.
\end{align}{}
Now, we can make the last three norms arbitrarily small and $b_1, b_2, b_3$ positive by choosing $k_\textup{c}$, $\sigma$ and $k_4$ small enough, we can make $b_4$ positive by choosing $(\sigma_i)_{\ i \in \mathcal{I}}$ small enough, we can make $b_5$ positive by making $k_{\textup{b}}$ large enough. Of the mentioned parameters, only  $k_\textup{b}$ and $k_\textup{c}$ cannot be chosen arbitrarily. To make $k_\textup{b}$ large enough, we chose $(K_i)_{\ i \in \mathcal{I}}$ and $(\rho_i)_{\ i \in \mathcal{I}}$ large enough, to make $k_\textup{c}$ small enough we have to chose parameters $k_1$ and $k_2$ small enough. Since $\Omega_c$ was chosen for arbitrary $c$, it follows that for any compact set $K$, it is possible to find such control parameters that for $({\bfs{\eta}}(0), \hat{\bfs{\theta}}(0), \bfs{u}(0)) \in K$, $(\tilde{\bfs{\eta}}, \tilde{\bfs{\theta}}, \bfs{u})$ converge to an arbitrarily small neighborhood of $(0, 0, \bfs{u}^*)$, which concludes the proof. For $\lambda$ we can only claim that this is bounded.

\section{Proof of Theorem 3}
%The full state dynamics consist of the GNE seeking scheme described in section 3 and the dynamics of the multi-agent system in \nref{eq: fast_dynamics}. As we have proven the stability of the GNE seeking scheme in the case of static systems and cost functions, and have assumed that a constant input exponential stabilizes the multi-agent system, 
We have to prove that there exists a timescale separation between the GNE learning scheme described in section 3 and dynamics of the multi-agent system in \nref{eq: fast_dynamics} such that the interconnection is also stable. Let us consider a Lyapunov function candidate $V = V_\theta + V_\omega + V_z$, where $V_\theta$ and $V_\omega$ are the same as \nref{eq: lyapunov parameter est}, \nref{eq: V_i} and $V_z$ is formed using Standing Assumption \ref{assum: lyap} in the following way: 
\begin{align}
    \textstyle V_z(\bfs{z}, \bfs{u}) = \sum_{i \in \mathcal{I}} V_i(z_i, u_i).\label{eq: lyapunov multiagent}
\end{align}{}
\emph{Outline of the proof:} We first bound all of the terms in $D^+ V_z$ introduced by nonconstant inputs with functions of the variables $(\bfs{\eta}, \hat{\bfs{\theta}}, \omega, \bfs{z})$, then we bound all of the terms in $D^+{V_\theta}$ introduced by the redefinition of error $e_i$ in \nref{def: new_error} in the same manner. At the end, we use the quadratic terms of the complete $D^+ V$ to show that the additional terms are majorized by the negative terms.\\ \\

\emph{Multi-agent term:} Let us do a change of variables $\bfs{z} = \bfs{x} - \pi(\bfs{u})$ in \nref{eq: fast_dynamics}. New dynamics read as
\begin{align}
    \epsilon \dot{\bfs{z}} = f(\bfs{z} + \pi(\bfs{u}), \bfs{u}) - \epsilon \nabla \pi(u) \dot{\bfs{u}}.\label{eq: fast dynamics z}
\end{align}{}
Dini derivative of \nref{eq: lyapunov multiagent}, by plugging in \nref{eq: fast dynamics z}, reads as
\begin{align}
    D^+ V_z(\bfs{z}, \bfs{u}) &= \nabla_{\bfs{z}}V_z^\top \dot{\bfs{z}} + \nabla_{\bfs{u}}V_z^\top \dot{\bfs{u}} \nonumber\\
    &= \frac{1}{\epsilon}\nabla_{\bfs{z}} V_z(\bfs{z}, \bfs{u})^\top f(\bfs{z} + \pi(\bfs{u})) \nonumber\\
    &\quad - \nabla_{\bfs{z}} V_z(\bfs{z}, \bfs{u})^\top  \nabla \pi(\bfs{u}) \dot{\bfs{u}} + \nabla_{\bfs{u}} V_z(\bfs{z}, \bfs{u})^\top \dot{\bfs{u}}\nonumber
\end{align}
By using Standing Assumption \ref{assum: lyap} and inequality \nref{eq: famous_ineq}, we can further improve the bound:
\begin{align}

     D^+ V_z(\bfs{z}, \bfs{u})&\leq -\frac{\kappa}{\epsilon}\|\bfs{z}\|^2 + L_{10} \|\bfs{z}\|\|\dot{\bfs{u}}\|\nonumber\\
    &\leq -\left(\frac{\kappa}{\epsilon} - \frac{L_{10} k_5}{2}\right)\|\bfs{z}\|^2 + \frac{L_{10}}{2 k_5} \|\dot{\bfs{u}}\|^2,\nonumber
\end{align}{}
where $L_{10} > 0$ is the Lipshitz constant of the function $\max_{\bfs{u} \in \mathcal{U}}\nabla_{\bfs{u}} V_z(z, \bfs{u}) - \nabla_{\bfs{z}} V_z(z, \bfs{u})^\top  \nabla \pi(\bfs{u})$ and $k_5 > 0$.\\ \\

\emph{Parameter estimation term:} GNE learning is identical as in the static case, apart from the measurements of the cost function. Let us denote
\begin{align}
    \bfs{l} &\coloneqq \col{\z{l_i}_{i \in \mathcal{I}}}, \nonumber \\
    \bfs{y} &\coloneqq \col{\z{y_i}_{i \in \mathcal{I}}}, \nonumber \\
    h(\boldsymbol{x})&\coloneqq \operatorname{col}\left(\left(h_i \left(u_{i}, \bfs{u}_{-i}\right)\right)_{i \in \mathcal{I}}\right).\nonumber
\end{align}{} The difference in the measurement introduces an additional component in the bound for the derivative of the Lyapunov function of the parameter estimation term:
\begin{align}
    &\| \tilde{\bfs{\eta}} \| \n{\dot{\bfs{y}} - \dot{\bfs{l}}} = \n{ \tilde{\bfs{\eta}} \|\| \tfrac{d}{dt}\z{{h}(\bfs{x}) - {h}(\pi(\bfs{u}))} } \nonumber \\
    &=  \| \tilde{\bfs{\eta}} \|\n{ \dfrac{1}{\epsilon} \nabla h(\bfs{x}) f(\bfs{x}, \bfs{u}) - \nabla h(\pi(\bfs{u})) \nabla \pi(\bfs{u}) \dot{\bfs{u}} } \nonumber\\
    &\leq \dfrac{1}{\epsilon}\| \tilde{\bfs{\eta}} \|\n{ \nabla h(\bfs{x}) f(\bfs{x}, \bfs{u}) - \nabla h(\bfs{x}) f(\pi(\bfs{u}), \bfs{u})} \nonumber \\
    &\quad +  \| \tilde{\bfs{\eta}} \|\n{\nabla h(\pi(\bfs{u})) \nabla \pi(\bfs{u}) \dot{\bfs{u}} } \nonumber\\
    &\leq \dfrac{L_{11}}{\epsilon}\| \tilde{\bfs{\eta}} \|\| \bfs{z}\| +  L_{12}\| \tilde{\bfs{\eta}} \|\|\dot{\bfs{u}} \|\nonumber\\
    &\leq \z{\dfrac{L_{11 k_6}}{2\epsilon} + \frac{L_{12} k_7}{2}}\| \tilde{\bfs{\eta}} \|^2 + \dfrac{L_{11}}{2\epsilon k_6}\| \bfs{z}\|^2 +  \frac{L_{12}}{2 k_7}\|\dot{\bfs{u}}\|^2,\nonumber
\end{align}{}
where $L_{11}, L_{12}, k_7, k_6 > 0$ and the second to last equation follows from the (local) Lipshitz continuity of the functions and the fact that the variables are bounded on a compact set $\Omega_c \coloneqq \{(\bfs{x}, \omega, \hat{\bfs{\eta}}, \hat{\bfs{\theta}}) \in \bfs{\mathcal{X} \times \mathcal{{U}}}  \times \R^{q} \times \R^N \times \Theta \ |\  V(\bfs{z}, \omega, \tilde{\bfs{\eta}}, \tilde{\bfs{\theta}}) \leq c\}$.\\ \\

\emph{Complete Lyapunov candidate:} Finally, the Dini derivative of the complete Lyapunov function candidate is bounded as follows:
\begin{align}
    &D^+{V_\theta}(\omega) + D^+{V_\theta}(\hat{\bfs{\eta}}, \hat{\bfs{\theta}}) + D^+ V_z(\bfs{z}, \bfs{u}) \nonumber\\ 
    &\leq  -\left(\frac{1}{2} - k_\textup{c} L_4 - (\sigma + k_\textup{c} L_6)L_1 - \frac{L_{10}}{2 k_5} - \frac{L_{12}}{2 k_7}\right)\| \dot{\bfs{u}}\|^2 \nonumber\\
    
    &-\left(\frac{1}{2} - k_\textup{c} L_5\right)\| \dot{\lambda}\|^2 - \left(k_{\textup{a}} - \dfrac{L_{11 k_6}}{2\epsilon} - \frac{L_{12} k_7}{2}\right)\|\tilde{\bfs{\eta}}\|^{2} \nonumber \\
    
    &- \left(k_{\textup{b}} - k_{\textup{c}} L_7 - \frac{\sigma_{\textup{max}}(\Gamma) k_3}{2}\right) \|\tilde{\bfs{\theta}}\|^{2} \nonumber\\
    
    &- \left(\frac{ \mu \sigma_{\textup{min}}}{2} - \frac{\sigma_{\textup{max}}(\Gamma)}{2 k_3} - \frac{1}{2 k_4} - (\sigma - k_\textup{c} L_6)L_2\right)\|\tilde{\bfs{u}}\|^2   \nonumber \\
    
    &-\left(\frac{\kappa}{\epsilon} - \frac{L_{10} k_5}{2} - \dfrac{L_{11}}{2\epsilon k_6}\right)\|\bfs{z}\|^2 - \left(\frac{1}{2}-k_\textup{c} L_8\right)\|\bfs{e} - \bfs{\bfs{\eta}}\|^2  \nonumber\\
    
    &+ k_\textup{c} L_9 \| \dot{d}(t) \|^2 + (\sigma + k_\textup{c} L_6)L_3 \|F(\bfs{u}^*) \|+ \frac{k_4}{2} \|d(t)\|^2\nonumber\\
    
    & \leqq -b_1 \| \dot{\bfs{u}}\|^2  -b_2 \| \dot{\lambda}\|^2 - b_3\|\tilde{\bfs{\eta}}\|^{2}  - b_4 \|\tilde{\bfs{\theta}}\|^{2}   - b_5 \|\tilde{\bfs{u}}\|^2  \nonumber\\
    
    &  - b_6 \|\bfs{z}\|^2 - b_7 \|\bfs{e} - \bfs{\bfs{\eta}}\|^2 +b_8\|F(\bfs{u}^*) \|^2 + b_9 \| \dot{d}(t) \|^2  \nonumber \\
    
    & + b_{10} \|d(t)\|^2.
\end{align}{}
% Now, we can make $k_\textup{a}$, $k_\textup{b}$ large enough by choosing $(K_i)_{\ i \in \mathcal{I}}$ and $(\rho_i)_{\ i \in \mathcal{I}}$ large enough, we can choose $k_5$ to $k_7$ large enough, we can make $k_\textup{c}$ small enough by choosing $k_1$ and $k_2$ small enough, and we can choose $(\sigma_i)_{\ i \in \mathcal{I}}$, $\epsilon$ small enough, such that all the norm squares are negative definite. It follows that $(\tilde{\eta}, \tilde{\theta}, \bfs{u}, \bfs{z})$ converge to a neighborhood of $(0, 0, \bfs{u}^*, 0)$, where the radius of the neighborhood is proportional to $\|d(t)\|^2, \|\dot{d}(t)\|^2$ and $\|F(u^*)\|^2$.

Now, we can make the $b_8, b_9, b_{10}$ arbitrarily small and $b_2, b_7$ positive by choosing $k_\textup{c}$, $\sigma$ and $k_4$ small enough, we can make $b_1$ positive by choosing $k_5, k_7$ large enough, we can make $b_3$ positive by choosing $k_\textup{a}$ large enough, we can make $b_4$ positive by choosing $k_\textup{b}$ large enough,  we can make $b_5$ positive by choosing $(\sigma_i)_{\ i \in \mathcal{I}}$ small enough and we can make $b_6$ positive by choosing $\epsilon$ small enough. Of the mentioned parameters, only  $k_\textup{a}$, $k_\textup{b}$ and $k_\textup{c}$ cannot be chosen arbitrarily. To make $k_\textup{a}$ and $k_\textup{b}$ large enough, we chose $(K_i)_{\ i \in \mathcal{I}}$ and $(\rho_i)_{\ i \in \mathcal{I}}$ large enough, to make $k_\textup{c}$ small enough we have to chose parameters $k_1$ and $k_2$ small enough. Since $\Omega_c$ was chosen for arbitrary $c$, it follows that for any compact set $K$, it is possible to find such control parameters that for $(\tilde{\bfs{\eta}}(0), \tilde{\bfs{\theta}}(0), \bfs{u}(0), \bfs{z}(0)) \in K$, $(\tilde{\bfs{\eta}}, \tilde{\bfs{\theta}}, \bfs{u}, \bfs{z})$ converge to an arbitrarily small neighborhood of $(0, 0, \bfs{u}^*, 0)$, which concludes the proof. For $\lambda$ we can only claim that this is bounded.\\ \\
\end{document}